\documentclass{article}                     
\usepackage{graphicx}
\usepackage{amsmath}
\usepackage[hyphens]{url}
\usepackage{natbib}
\usepackage{multirow}
\usepackage{empheq}

\graphicspath{{./}}  

\begin{document}

\title{On self-organised aggregation dynamics in swarms of robots with informed robots
}


\author{Ziya Firat \\Faculty of Computer Science, University of Namur, Namur, Belgium\\ ziya.firat@student.unamur.be\and
        Eliseo Ferrante\\ University of Birminghaam, Dubai, UAE\\ e.ferrante@bham.ac.uk \and
        Yannick Gillet\\
        Faculty of Computer Science, University of Namur, Namur, Belgium\\ yannick.gillet@unamur.be\and
        Elio Tuci\\ Faculty of Computer Science, University of Namur, Namur, Belgium\\ elio.tuci@unamur.be
}




\maketitle

\begin{abstract}
	Aggregation is a process observed in natural systems whereby individuals gather together to form large cluster. Recent studies with cockroaches and robots have shown that relatively simple individual mechanisms can account for how individuals manage to gather on a single shelter when two or more are available in the environment. In this paper, we use simulated swarms of robots to further explore the aggregation dynamics generated by these simple individual mechanisms. Our objective is to study the introduction of ``informed robots'', and to study how many of these are needed to direct the aggregation process toward a pre-defined site among those available in the environment. Informed robots are members of a group that selectively avoid the site/s where no aggregate should emerge, and stop only on the experimenter predefined site/s for aggregation. We study the aggregation process with informed robots in three different scenarios: two that are morphologically symmetric, whereby the different types of  aggregation site are equally represented in the environment; and an asymmetric scenario, whereby the target site has an area that is half the area of the sites that should be avoided. We first show what happens when no robot in the swarm is informed: in symmetric environments, the swarm is able to break the symmetry and aggregates on one of the two types of site at random, not necessarily on the target site, while in the asymmetric environment, the swarm tends to aggregate on the sites that are most represented in terms of area. The original contribution of this study is to demonstrate the effect of the introduction of a small proportion of informed robots in both environments: In symmetric environments, they selectively direct the aggregation process towards the experimenter chosen site; in the asymmetric environment, informed robots can invert the spontaneous preference for the most represented site and induce the swarm to aggregate on the least represented type of site. Moreover, for each scenario, we analyse how the dynamics of the aggregation process depends on the proportion of informed robots. As a further valuable contribution of this study, we provide analytical results by studying a system of Ordinary Differential Equations' (ODEs) that is an extension of a well known model. Using this model, we show how, for certain values of the  parameters, the model can predict the dynamics observed with simulated robots in one of the two symmetric scenarios. 
\end{abstract}

\section{Introduction}
\label{sec::intro}
The aim of this study it to illustrate how a swarm of autonomous mobile robots can be induced to aggregate on a desired aggregation site chosen by the experimenter among those available in the robots' environment, simply by informing a small fraction of the swarm about which is the target site. This study extends and deepens our understanding of the aggregation dynamics in swarm robotics, by showing that ``informed robots'' is a  relatively simple and rather effective means to control the swarm.

Swarm robotics is a sub-domain of a larger research area dedicated to the design and control of multi-robot systems~\citep[see][]{sahin2004swarm,brambilla2013swarm}. Swarm robotics is characterised by the following distinctive elements: i) the use of distributed embodied control, that is, each robot has is own on-board control system in charge of determining the robot's behaviour; ii) local perception, that is, each robot can sense and communicate only within a given range using sensors and actuators mounted on its body; iii) the use of indirect communication: given that the robots of a swarm are ``anonymous'', any single agent can not selectively choose a specific message receiver, but rather communicate implicitly through modification of the environment in which they operate. The latter can be done by emitting sound or by generating other types of signal that are eventually detected by other agents.

Research in swarm robotics generally focuses on the design of individual control mechanisms underpinning a desired collective response, which emerges in a self-organised way from the interactions of system components~\citep[i.e., the robots and their environment, see][]{brambilla2013swarm}. Examples of such collective responses are area
coverage~\citep[][]{HauertEtAl2008}, chain formation~\citep[][]{SperatiEtAl2011}, collective decision-making and task partitioning~\citep[][]{MontesDeOcaEtAl2011,PiniEtAl2011,TuciEtAl2015}, cooperative transport~\citep[][]{AlkilabiEtAl2017}, and collective motion~\citep[][]{FerranteSACOMM2011}. 

One of the main building blocks in swarm robotics is collective decision-making; that is, the ability to make a collective decision without any centralised leadership, but only via local interaction and communication. Several types of collective behaviours can be seen as instances of collective decision making~\citep[see][]{Valentini2017Review,ValentiniEtAL2015}, including collective motion where robots have to agree on a common direction of motion, and aggregation where robots have to agree on a common location in the environment~\citep[see][]{garnier2005agg,garnier2008agg,Bayindir:2009,correll:2011,gauci2014self}. Indeed, aggregation is often a necessity for many swarm robotic systems as it is a prerequisite for other cooperative behaviours~\citep[][]{dorigo2004evolving,TuciEtAl2018}.

Aggregation processes are extremely common in biological systems, resulting in clusters of agents at common areas in the environment~\citep[][]{camazine2003self}. Self-organised aggregation (i.e., an aggregation process not driven by exogenous forces) has been studied in a variety of biological systems~\citep[][]{deneubourg2002dynamics,Jeanson:2005}. In a seminal work illustrated in~\citep[][]{AmeEtAl2004_AB}, the authors describes a mathematical model that accounts for the aggregation behaviour observed in cockroaches by linking the individual resting time to the perception of conspecific resting on the aggregation site. Generally speaking, the model provides a rational to why individuals of different strains aggregate on a single resting site in spite of segregation dynamics induced by chemical signals that would tend to generate same-strain individuals aggregates. In~\citep[][]{AmeEtAlPNAS2006}, the authors extend the above mentioned mathematical model predicting that, in an environments with up to four aggregation sites, cockroaches form a single aggregate only when each aggregation site can host more that the totality of the individuals. The model also predicts how the cockroaches distribute in different environment varying for the number of aggregation sites and the diameter of each site bearing upon the site capacity to host individuals.

The principle of attraction between individuals, that nicely accounts for the aggregation dynamics observed in cockroaches, has been ``imported'' into robotics to design effective and relatively simple control mechanisms to achieve aggregation behaviour. In particular, roboticists have shown that robot's controllers in which the individual probability to join and to leave an aggregation site depend on the number of robots perceived by an individual at the site, lead to the formation of a single aggregate in environments with multiple aggregation sites. Both in~\citep[][]{GarnierEtAl2009} and in~\citep[][]{CampoEtAl2010} the authors have used the above mentioned probabilistic controller to look at aggregation dynamics in scenarios with two circular aggregation sites; that is, sites that the robots can perceive but can not distinguish one from the other. In~\citep[][]{GarnierEtAl2009}, the authors considered two cases: in the first one, where the two sites have equal size, under special circumstances the swarm can break the symmetry and aggregate on one of the two sites at random; in the second one, where the two sites have different sizes, robots are able to collectively chose the biggest among the two aggregates. In~\citep[][]{CampoEtAl2010}, the robots are required to aggregate in environments in which the carrying capacity of a site is systematically varied while the carrying capacity of the target site remains fixed to a value that allows the site to host all the robots of the swarm. The study shows that robots avoid aggregation sites that are too small or too big with respect to the swarm size.

In this paper, we explore the aggregation dynamics generated by the principle of attraction between individuals in a novel setting. We introduce the concept of ''informed robots'', robots that are apriori informed on which site they need to stop. Apart from the additional capacity to avoid to stop on undesired aggregation sites, informed robots behave exactly like any other robot in the swarm. As designers, we hypothesise that the effect of informed robots is to stir the aggregation dynamics towards a specific site chosen by the experimenter among those available to the robots. The effects of informed individuals on groups dynamics has been recently investigate in biology to account for the motion of collective systems, such as birds and fishes. In a seminal study illustrated in~\citep[][]{CouzinEtAl2005}, the authors study collective decision making in the context of collective motion looking at what happens when \emph{implicit leaders} are introduced. These special individuals, also called informed individuals, have a preferred direction of motion and they bias the collective decision in that direction. The rest of the swarm does not have any preferred direction of motion, nor is able to recognise informed individuals as such. The authors show that the accuracy of the group motion towards the direction known by the informed agents increases asymptotically as the proportion of informed individuals increases. Moreover, the authors show that larger the group, the smaller the proportion of informed individuals needed to guide the group with a given accuracy. 

In swarm robotics, the framework of implicit leaders has been studied mainly in the context of collective motion~\citep[][]{Celikkanat2010,FerranteNOALIGN,FerranteSACOMM2011}. Inspired by these works, we study the effect of implicit leaders in the context of self-organised aggregation. Differently from the aggregation studies mentioned above, and analogously to the studies performed within collective motion~\citep[][]{CouzinEtAl2005}, in this paper we introduce informed robots in the context of self-organised aggregation, and we study how they impact the aggregation dynamics. 
The roles of informed robots is to influence the aggregation dynamics, in a very indirect way, since none of the robots has any means to discriminate informed from non-informed robots.

We perform our study with a series of simulation experiments on different scenarios, represented by a circular arena with two to four aggregation sites. In the simplest possible scenario, the arena is characterised by two aggregation sites, the desired one coloured in black, and the one to be avoided, coloured in white. Only informed robots are programmed to avoid to stop on the white site. For all the other robots of the swarm, both the black and the white site are equally good resting locations.  We show that with less than 20\% of informed robots, the swarm systematically aggregate on the desired black site (see section~\ref{sec::results:2_sites}, Exp. I). The results of subsequent experiments show that with a slightly larger proportion of informed robots the swarm can systematically aggregate on the desired black site even in an ``asymmetrical'' scenario with two white and only one black site, and also in a scenario with two white and two black sites (see section~\ref{sec::results:3_4_sites}, Exp. II, and Exp. III). For each experiment, we provide an analysis of the aggregation dynamics that lead to the formation of a single aggregate. We also show interesting relationships between swarm size and proportion of informed agents, both on quality and speed of convergence on the desired aggregation site. Moreover, we propose an Ordinary Differential Equation model that extends the one originally illustrated in~\citep[][]{AmeEtAlPNAS2006} in a way to include the effect of informed robots. The analysis of the model with the same parameters as in~\citep[][]{AmeEtAlPNAS2006} indicates that a very large proportion of informed robots (i.e., about 80\%) is needed to qualitatively replicate the aggregation dynamics we observed in the  two-site scenario. However, by exploring the parameter space of the model in a way that goes beyond the analysis done in~\citep[][]{AmeEtAlPNAS2006}, we identified a region of the parameter space whereby the stable equilibria qualitatively match those found in experiments with simulated robots. The analysis of the model's parameters leads to a deeper understanding of the relationships between environmental features and agents' exploration strategies. We show how these relationships bear upon the emergence of a single aggregate and how they interfere by amplifying or by reducing the effects of informed robots on the group aggregation process. 

The rest of the paper is structured as follows. Section~\ref{sec::fsm} describes the self-organised aggregation method used. In Section~\ref{sec::expsetup}, we present the experimental setup and how we study the effect of informed robots. Section~\ref{sec::results} presents the results of the three experiments with simulated robots. Section~\ref{sec::ode_model} shows the analysis of the ODEs' model. Finally, in section~\ref{sec::conclusions}, we discuss the significance of our results for the swarm robotics community, and we point to interesting future directions of work.

\section{The robots' controller}
\label{sec::fsm}
\begin{figure}[t]
	\begin{center}
			\includegraphics[width=0.3\textwidth]{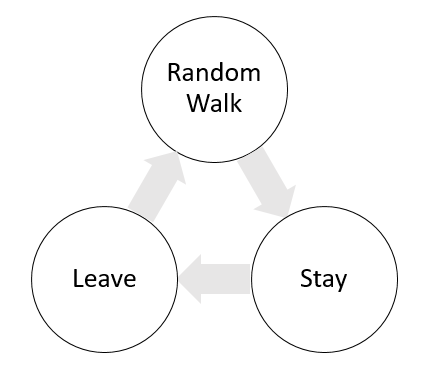}
		\caption{State diagram of the robots' controller.}
		\label{fig::controller}
	\end{center}
\end{figure}
Each robot is controlled by a probabilistic finite state machine (PFSM, see also Figure~\ref{fig::controller}), similar to the one employed in~\citep[][]{Jeanson:2005,Bayindir:2009,correll:2011,Cambier:2018}. The PFSM is made of three states: Random Walk ($\mathcal{RW}$), Stay ($\mathcal{S}$), and Leave ($\mathcal{L}$). When in state $\mathcal{RW}$, the robot performs a random walk strategy first introduced in~\citep{} that is very effective in covering the whole environment: Using this strategy, the movement of the robot is characterised by an isotropic random walk, with a fixed step length (5 seconds, at 10 cm/s), and turning angles chosen from a wrapped Cauchy probability distribution characterised by the following PDF~\citep[][]{kato2013}:
\begin{align}
f(\theta, \mu, \rho)=\frac{1}{2\pi}\frac{1-\rho^{2}}{1+\rho^{2}-2\rho\cos(\theta - \mu)}, \; 0 < \rho < 1,
\end{align}
where $\mu=0$ is the average value of the distribution, and $\rho$ determines the distribution skewness. For $\rho = 0$ the distribution becomes uniform and provides no correlation between consecutive movements, while for $\rho = 1$ a Dirac distribution is obtained, corresponding to straight-line motion. In this study $\rho = 0.5$. Any robot in state $\mathcal{RW}$ is continuously performing an obstacle avoidance behaviour. To perform obstacle avoidance, first the robot stops, and then it keeps on changing its headings of a randomly chosen angle uniformly drawn in $[0, \pi]$ until no obstacles are perceived.

A robot that, while performing random walk, reaches an aggregation site, it stops with probability ($P_{stay}$). This probability is computed using the following function:
\begin{equation} 
P_{stay} = 0.03 + 0.48 * (1 -  e^{-an});\label{eq::p_staty}
\end{equation}
with $n$ corresponding to the number of other robots currently stationing on the site that are perceived by the robot currently deciding whether to stop or not; and $a=2.6$. This function was first introduced in~\citep[][]{Cambier:2018}. It interpolates the probability table considered in classical studies such as~\citep[][]{Jeanson:2005,correll:2011}. Once the robot has decided to stop based on $P_{stay}$, it moves forward for a limited number of time in order to avoid stopping at the border of the site thus creating barriers preventing the entrance to other robots, and at the same time attempting to distribute uniformly with other robots on the site. It then transitions from state $\mathcal{RW}$ to state $\mathcal{S}$. Once in state $\mathcal{S}$ the robot leaves the aggregation site with probability $P_{leave}$. This probability is computed in the following:
\begin{equation} 
P_{leave} = e^{-bn};\label{eq::p_leave}
\end{equation}
with $b=2.2$. This function was also introduced in~\citep[][]{Cambier:2018}. A robot that decides to leave the aggregation site based on $P_{leave}$ transitions from state $\mathcal{S}$ to state $\mathcal{L}$. Both $P_{stay}$ and $P_{leave}$ are sampled every 20 time steps. When in state $\mathcal{L}$, the robot moves away from the site by moving forward while avoiding collisions with other robots until it no longer perceives the site. At this point, the robot transitions from state $\mathcal{L}$ to state $\mathcal{RW}$.

In our model we consider two kinds of robots: \emph{informed} and \emph{non-informed}. Informed robots are agents that possess extra information on what is the site or the sites on which the swarm has to aggregate. Ideally, this extra information could be either generated by additional sensors, mounted only on informed robots, which allow these robots to perceive the quality difference among the available aggregation sites, or communicated by the experimenter with the intention to influence the swarm aggregation dynamics. In our simulated scenario, we consider aggregation sites in two different colours: black and white. Informed robots are aware that the task requires to stop only on black sites. This information is implemented into the PFSM of informed robots with the instruction: do not stop on any white site. This means that informed robots ignore white sites, and only stop on black sites based on $P_{stay}$, as described above. Non-informed robots do not possess this extra information, therefore they can stop both on black and on white sites based on $P_{stay}$, as described above. Recall that any robot is not able to recognise whether any other individual is an informed or a non-informed robots.

\section{Experimental Setup}
\label{sec::expsetup}
\begin{figure}[t]
	\begin{center}
		\begin{tabular}{cc}\\
			\includegraphics[width=0.2\textwidth]{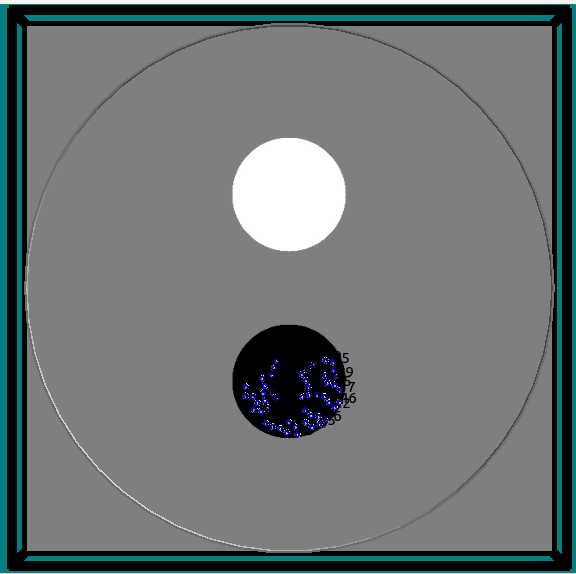}&
			\includegraphics[width=0.2\textwidth]{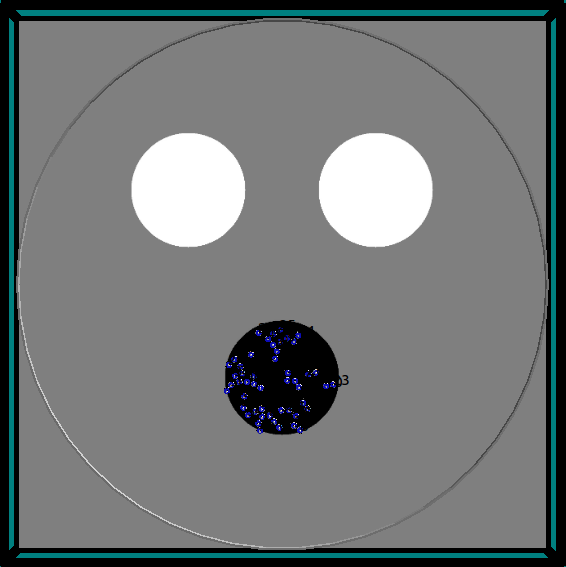}\\
			(a) & (b) \\
			\multicolumn{2}{c}{\includegraphics[width=0.2\textwidth]{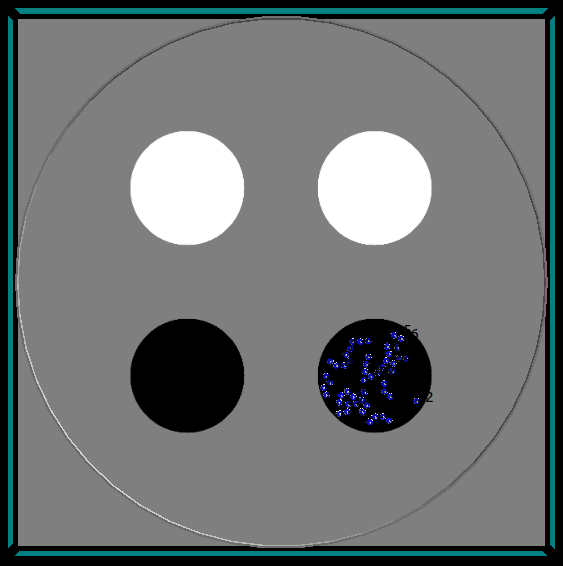}}\\
			\multicolumn{2}{c}{(c)}\\
		\end{tabular}
		\caption{The robots' arena for (a) experiment I, with two aggregation sites, one black and one white; (b) experiment II, with three aggregation sites, two white and one black; (c) experiment III, with four aggregation sites, two white and two black.}
	\label{fig::arena}
	\end{center}
\end{figure}
\begin{table}[b]
	\caption{Table showing, for each experiment (Exp.), the characteristics of each experimental condition. In Exp. I, there are two aggregation sites  in the arena (one black and one white); in Exp. II, there are three aggregation sites in the arena (one black and two white); in Exp. III, there are four aggregation sites in the arena (two black and two white).}
	\label{tab::sizes}
	\centering
	\begin{tabular}{|c|c|c|c|}\hline
		Exp. & \parbox{1.5cm}{\centering swarm\\ size (N)} & \parbox{1.5cm}{\centering arena\\ radius (m)} & \parbox{2.5cm}{\centering aggregation site\\ radius (m)} \\\hline
		\multirow{3}{*}{I} & 20 & 4 & 0.9 \\\cline{2-4}
		 & 50 & 6.32 & 1.4 \\\cline{2-4}
		 & 100 & 8.94 & 2.0 \\\hline

		\multirow{3}{*}{II} & 20 & 4.17 & 0.9 \\\cline{2-4}
		& 50 & 6.47 & 1.4 \\\cline{2-4}
		& 100 & 9.16 & 2.0 \\\hline
		
		\multirow{3}{*}{III} & 20 & 4.19 & 0.9 \\\cline{2-4}
		& 50 & 6.62 & 1.4 \\\cline{2-4}
		& 100 & 9.38 & 2.0 \\\hline
		
	\end{tabular}
\end{table}
In this set of simulations, a swarm of robots is randomly initialised in a circular area with the floor coloured in grey except for the circular aggregation sites, where the floor can be coloured in white or in black (see Figure~\ref{fig::arena}). We have studied three different scenarios. In experiment Exp. I, there are two aggregation sites in the arena, one black and one white. In experiment Exp. II, there are three aggregation sites in the arena, one black and two white. In experiment Exp. III, there are four aggregation sites in the arena, two black and two white. The task of the robots is to find and aggregate on a single black site. Some of the robots are informed on which type of site (i.e., black or white) to aggregate. The proportion of informed robots, henceforth denoted as  $\rho_I$ is systematically varied from $\rho_I=0$ (i.e., no robot is informed on which type of site to aggregate) to $\rho_I=1$ (all the robots are informed on which type of site to aggregate) with a step size of $0.1$. For each experiment, we have three different conditions, in which we varied the swarm size (N). As aggregation performance are heavily influenced by swarm density~\citep[][]{Cambier:2018}, in this paper we have decided to study scalability with respect to the swarm size by keeping the swarm density constant. Therefore, the diameter of the area as well as the diameters of the aggregation sites are varied in proportion to N. Table~\ref{tab::sizes} reports a summary of all experimental conditions. Note that in all experimental conditions, the area of each aggregation site is always large enough to accommodate all the robots of the swarm.

For each Exp., each condition can be divided in 11 tests which differ in the proportion of informed robots $\rho_I$. For each test, we execute 200 independent runs. In each run, the robots are randomly initialised within the arena, and then they are left free to act according to actions determined by their PFSM for 100.000 time steps. One simulated second corresponds to 10 time steps. Each run differs from the others in the initialisation of the random number generator, which influences all the randomly defined features of the environment, such as the robots initial position and orientation, as well as noise added to sensors readings.

We use ARGoS multi engine simulator~\citep[see][]{Pinciroli:SI2012}. The simulation environment models the circular arena as detailed above, and the kinematic and sensors readings of the Foot-bots mobile robots~\citep[see][]{bonani2010marxbot}. The robot sensory apparatus includes the proximity sensors positioned around the robot circular body, four ground sensors positioned two on the front and two on the back of the robot underside, and the range and bearing sensor. The proximity sensors are used for sensing and avoiding the walls of the arena. The readings of each ground sensors is set to $0.5$ if the sensor is on grey, to $1$ if on white, and $0$ if on black. A robot perceives an aggregation site when all the four ground sensors return a value different from $0.5$. Range and bearing sensors are used for inter-robot obstacle avoidance and for sensing the number of neighbours: the robots send a signal whenever they are stationing on a site. These signals are used by the robots to estimate the parameter $n$ necessary to compute $P_{stay}$ and $P_{leave}$. The maximum number of neighbours a robot can detect is $12$.

\section{Results}
\label{sec::results}
\begin{figure}[t]
	\begin{center}
		\begin{tabular}{cc}
			\includegraphics[width=0.28\textwidth]{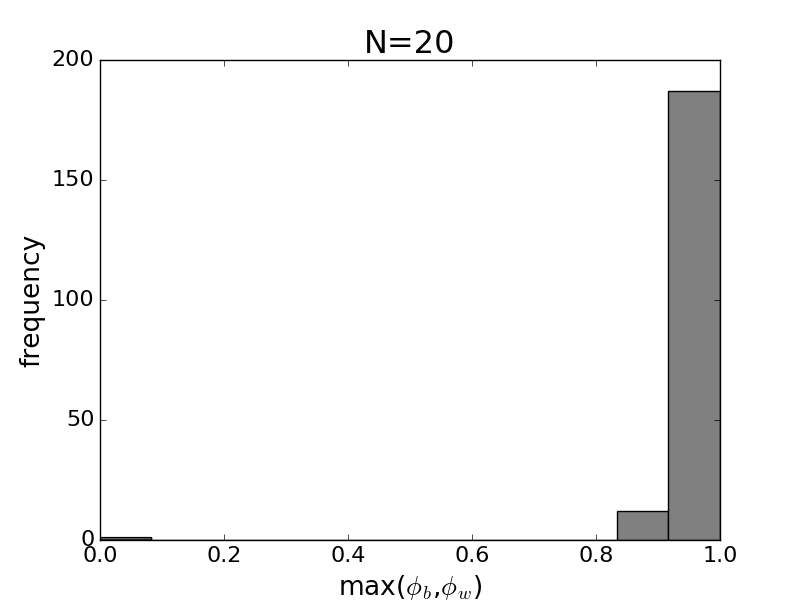} &
			\includegraphics[width=0.4\textwidth]{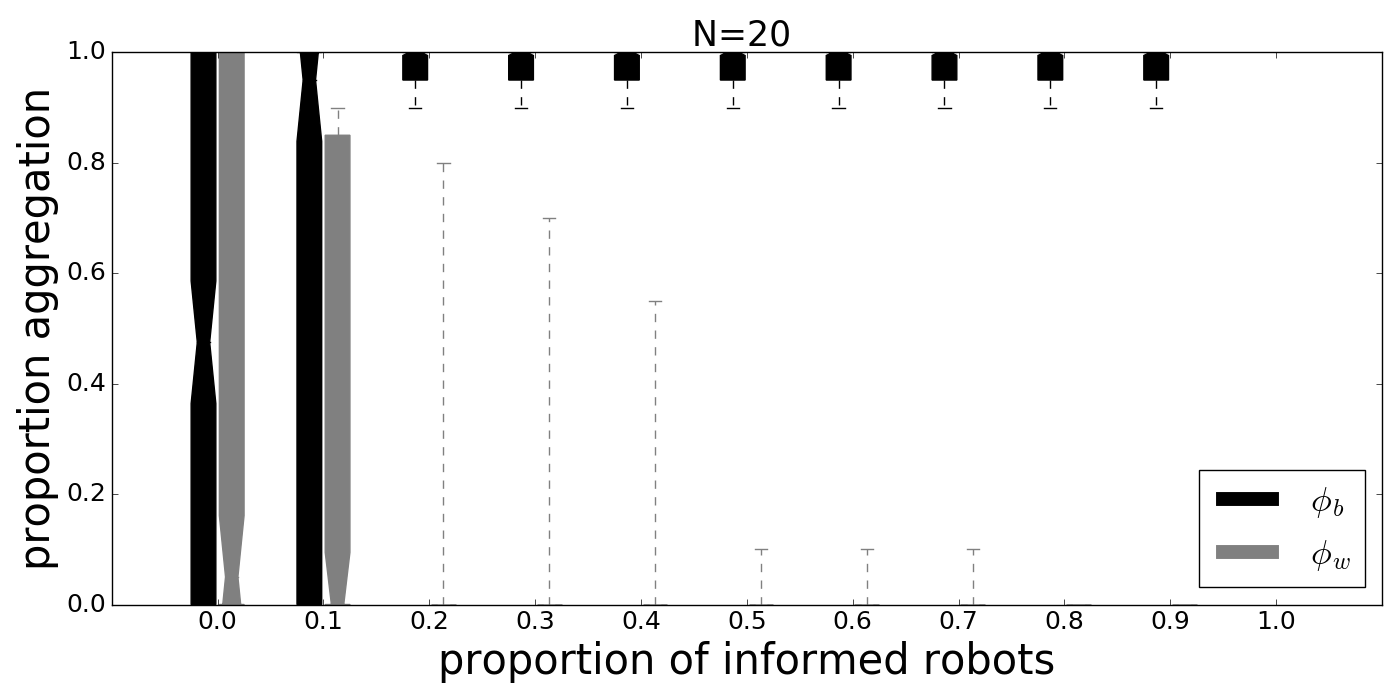} \\
			(a) & (b) \\
			\includegraphics[width=0.28\textwidth]{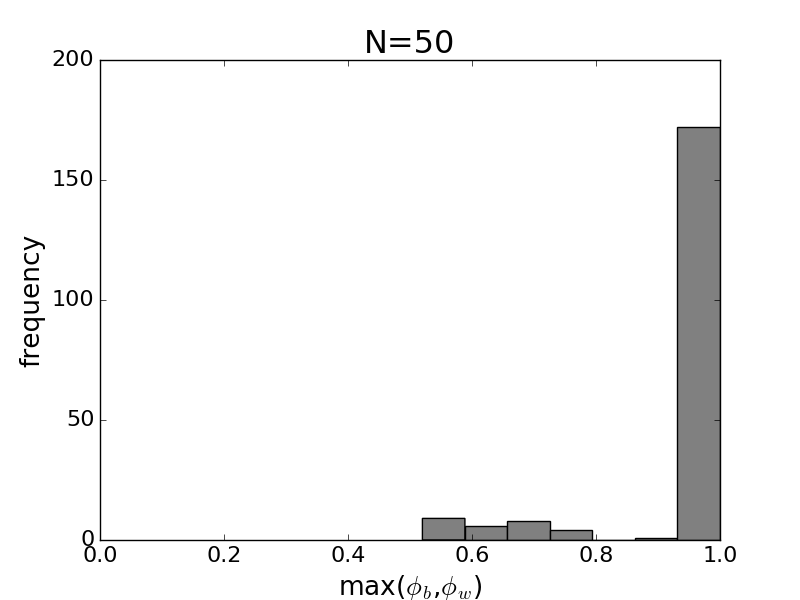}&
			\includegraphics[width=0.4\textwidth]{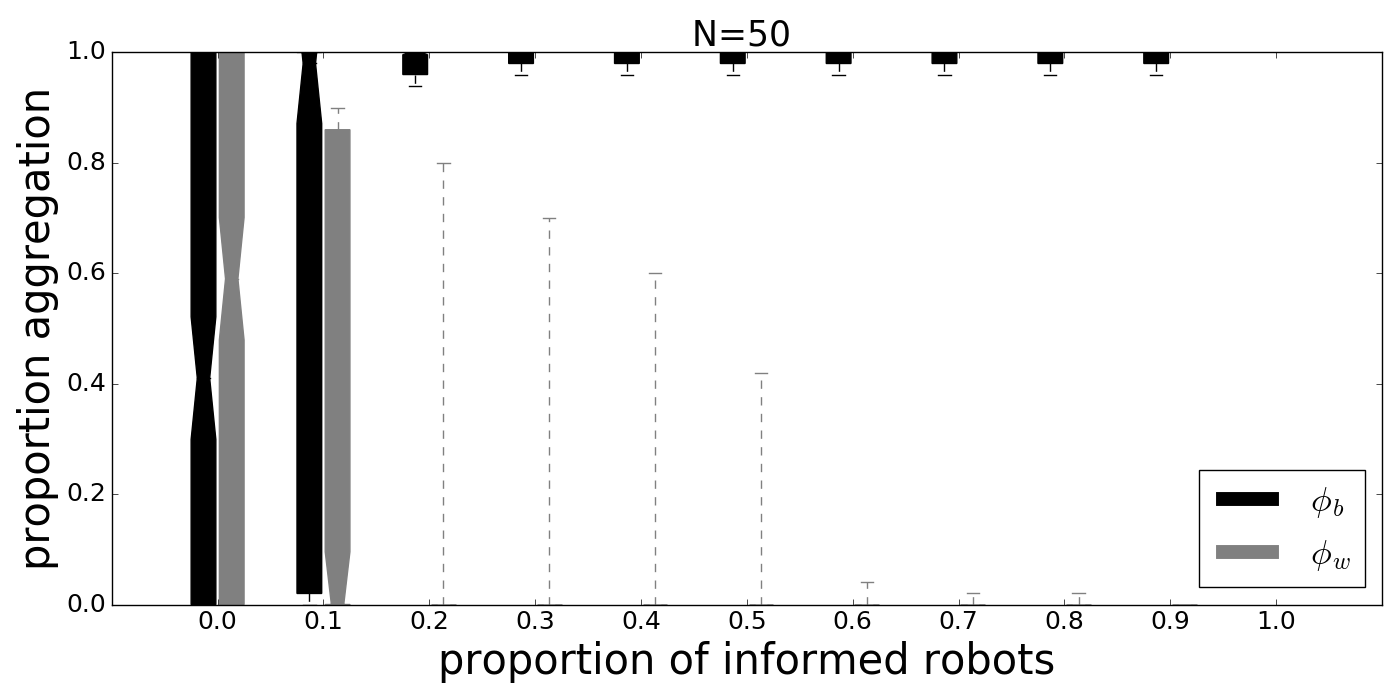}\\			
			(c) & (d) \\
			\includegraphics[width=0.28\textwidth]{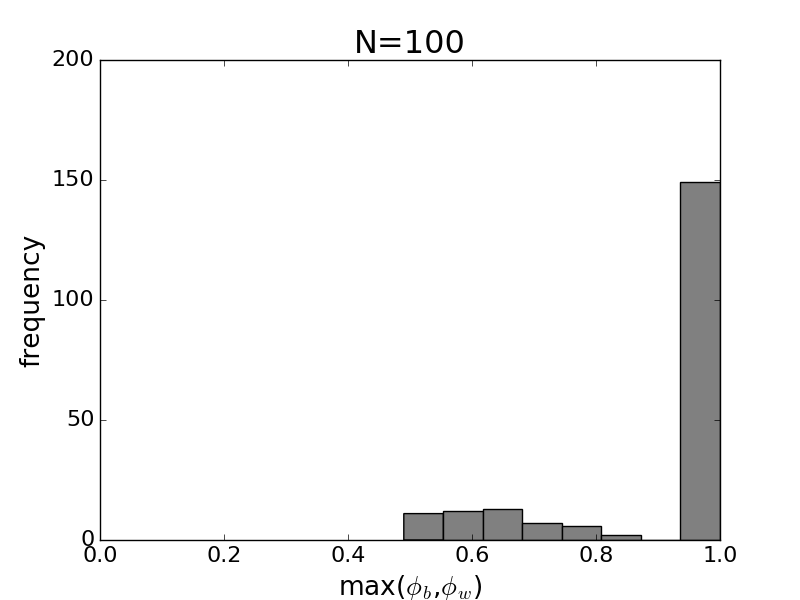}&
			\includegraphics[width=0.4\textwidth]{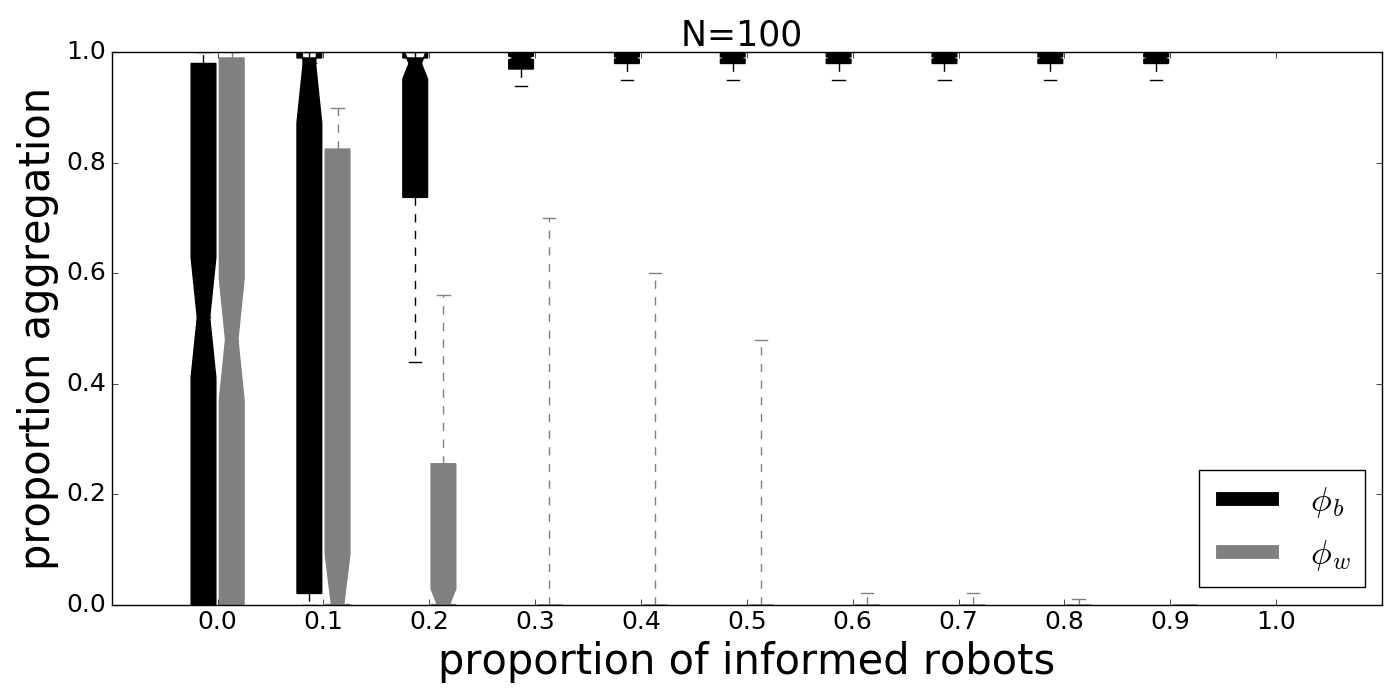}\\
			(e) & (f) \\
		\end{tabular}
		\caption{Results of the experiments in arena with two aggregation sites. The graphs in first column show frequency histograms of the proportion of robots aggregating on the largest aggregate ($\max(\Phi_b,\Phi_w)$) for swarms without informed robots ($\rho_I=0$) of size a) N=20; c) N=50; and e) N=100. Graphs on the second column show the percentage of aggregated robots on the white site (i.e., $\Phi_w$, see grey boxes) and on the black site ($\Phi_b$, see black boxes) for swarms of size b) N=20; d) N=50; and f) N=100. In each graph, the x-axis refers to the proportion of informed robots. Each box is made of 200 observations. Every observation in black/white boxes corresponds to $\Phi_w$/$\Phi_b$ computed at the end of a single run, respectively. Boxes represent the inter-quartile range of the data, while the thinnest point marks the median values. The whiskers extend to the most extreme data points within 1.5 times the inter-quartile range from the box.}
		\label{fig::2_sites}
	\end{center}
\end{figure}
The main aim of this study is to look at how informed robots influence the aggregation dynamics in scenarios where there are two or more possible aggregation sites, that can be differentiated only by informed robots. To do this, we used as performance indicator the proportion of robots aggregated on black site as $\Phi_b = \frac{N_b}{N}$ and on white site as $\Phi_w= \frac{N_w}{N}$ (where $N_b$ and $N_w$ are the number of robots aggregated on the black and white site, respectively) at the end of each run (i.e., after 100.000 time steps). For those scenarios with more than one site of the same colour as in Exp. II and in Exp. III, $N_w$ refers to the largest aggregate on white sites while $N_b$ refers to the largest aggregate on black sites. The goal of the swarm is to maximise $\Phi_b$ and to minimise $\Phi_w$. Note that $\Phi_b + \Phi_w \leq 1$ as it is possible that not all robots have aggregated in either type of site by the end of a run.

\subsection{Exp. I: arena with two aggregation sites, one black and one white}
\label{sec::results:2_sites}
In this section, we describe the results of Exp. I, which refers to the two-site aggregation scenario, with one black and one white site (see Figure~\ref{fig::arena}a). Prior to testing the effect of informed robots, we conduct a first set of experiments to validate our model. The model we used is strongly influenced by the work of~\citet[][]{GarnierEtAl2009}. According to this study, in presence of perfectly symmetrical aggregation sites, this aggregation model results in a symmetry breaking, whereby robots tend to chose one of the two sites at random. They aggregate in the chosen site, provided that the site is big enough to host the entire swarm. This symmetry breaking property is an essential feature of a self-organised aggregation method as it provides the positive feedback mechanism necessary for such behaviour. In order to test whether our model has this feature, we have executed experiments without informed robots in order to replicate the results in~\citep[][]{GarnierEtAl2009}. To calculate the strength of the positive feedback mechanism, we calculate the proportion of robots aggregated in the largest aggregate as $\max(\Phi_b,\Phi_w)$, independently on whether it is on the black or the white site. Results are shown in Figure~\ref{fig::2_sites} first column, in form of frequency distribution. The graphs shows that, independently on the swarm size, the distribution looks multi-modal, with the highest peak at $1.0$. This indicates that, for all considered swarm sizes, the swarm is able to create large aggregates (i.e., larger than 90\% fo the swarm size) around one of the sites.


With the introduction of informed robots, we analyse how aggregation performance depend on their proportion (i.e., $\rho_I$). The results are shown in Figure~\ref{fig::2_sites} second column. We notice that for all swarm size, and when no robot is informed in the swarm ($\rho_I=0.0$), both $\Phi_b$ and $\Phi_w$ are centred around $0.5$ with a strong variation. Without informed robots, aggregates that include more than 90\% of the swarm's components occur 98 times on the white site and 100 times on the black site in 200 runs for N=20;  94 times on the white site and 79 times on the black site in 200 runs for N=50; 72 times on the white site and 77 times on the black site in 200 runs for N=100. In summary, without informed robots, swarms form relatively large (i.e, with more than 90\% of the swarm's components) aggregates quite frequently (99\% of the runs for N=20, 86\% of the runs with N=50, and 75\% of the runs with N=100). These aggregates can be either on the black or on the white site. This may be explained by the fact that robots chose one aggregate at random without informed robots. 

The introduction of as little as 10\% of informed robots clearly breaks the almost equal-frequency bimodal aggregation dynamics between the black and the white site and generates new dynamics that tend to bring the majority of the robots on the black site. Furthermore, all three graphs in Figure~\ref{fig::2_sites} second column, show a similar trend in which the higher the number of informed robots, the higher the proportion of robots aggregated on the black site. This trend is non-linear and reaches saturation at around $\rho_I=0.2$ for $N=20$ and $N=50$, and for $\rho_I=0.3$ for $N=100$. With as little as 20\% to 30\% of informed robots, the totality of the runs finishes with more than 90\% of the robots aggregated on the black site (see black boxes for $\rho_I=0.2$ in Figure~\ref{fig::2_sites}b and~\ref{fig::2_sites}d, and for $\rho_I=0.3$ in Figure~\ref{fig::2_sites}f). For the smallest and the medium swarm size (N=20 and N=50, see Figure~\ref{fig::2_sites}b and~\ref{fig::2_sites}d) 20\% of informed robots is enough to bring forth very robust and consistent aggregation dynamics that take the entire swarm on the black site. For the largest swarm size, similar robust and consistent dynamics can be observed when the proportion of informed robots is at least 30\% (N=100, see Figure~\ref{fig::2_sites}f). In summary, the above results indicate that with a proportion of informed robots varying from $0.2$ to $0.3$ of the entire swarm, it is possible to generate robust and consistent aggregation dynamics that take the totally of the swarm on a single site, in a task in which two possible aggregation sites are available.

\begin{figure}[t]
	\begin{center}
		\begin{tabular}{ccc}
			\includegraphics[width=0.33\textwidth]{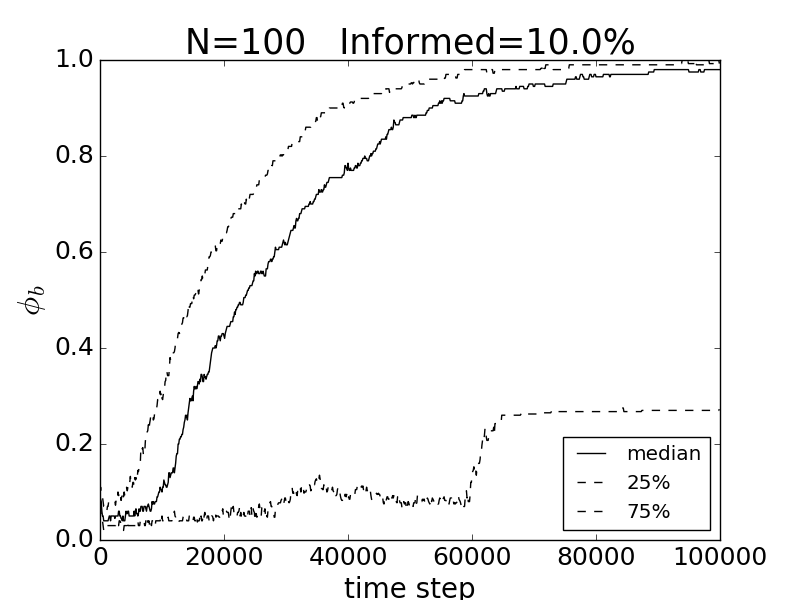} &
			\includegraphics[width=0.33\textwidth]{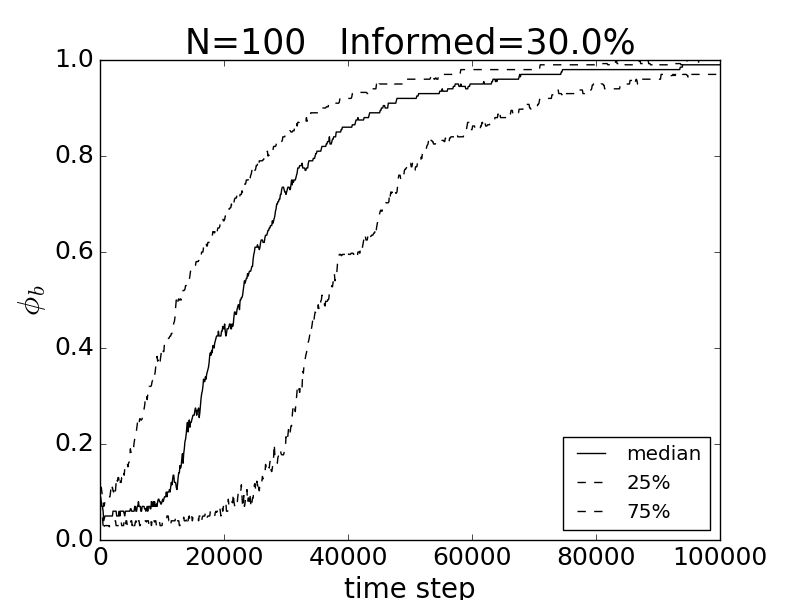} &
			\includegraphics[width=0.33\textwidth]{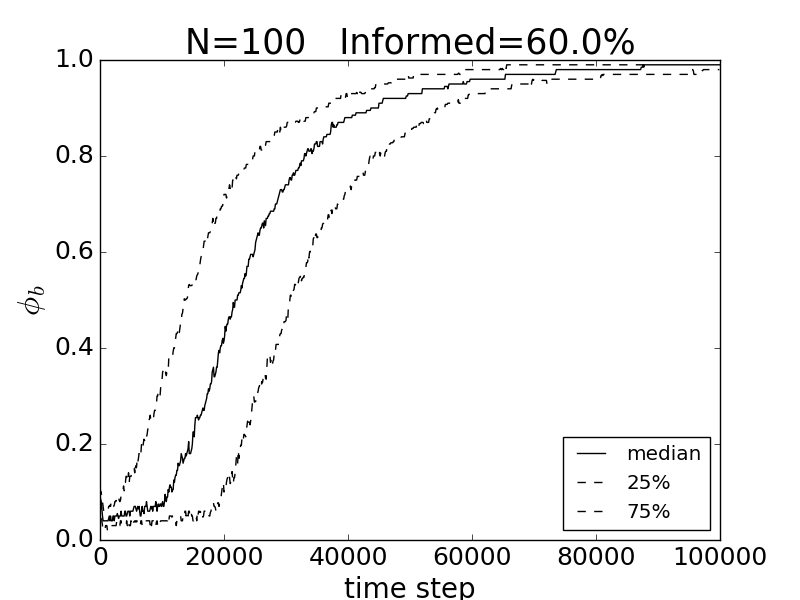} \\
			(a) & (b) & (c) \\
		\end{tabular}
		\caption{Graphs showing the median (see continuous line), the first and third quartile (see dashed lines) of the proportion of robots aggregated on the black site ($\Phi_b$) at every time step for 200 runs with swarm size N=100, in arena with two aggregation sites. In a) 10\% of the swarm is informed; in b) 30\% of the swarm is informed; in c) 60\% of the swarm is informed.}
		\label{fig:results:time_evol:2_sites}
	\end{center}
\end{figure}
The graphs in Figure~\ref{fig:results:time_evol:2_sites} show details on the time dynamics of the aggregation process for three different values of $\rho_I$ ($\rho_I=0.1$ in Figure~\ref{fig:results:time_evol:2_sites}a, $\rho_I=0.3$ in Figure~\ref{fig:results:time_evol:2_sites}b, and $\rho_I=0.6$ in  Figure~\ref{fig:results:time_evol:2_sites}c) and with the largest swarm size $N$=100. All figures feature a non-linear increase of the proportion of robots aggregated on the black site (i.e., $\Phi_b$), which eventually reaches saturation. By increasing the percentage of informed robots, we initially observe that distribution of convergence values changes dramatically from $\rho_I=0.1$ to $\rho_I=0.3$. In the latter case, we already observe the almost totality of the runs converging to all robots aggregated on the black site, as the dashed top curve in Figure~\ref{fig:results:time_evol:2_sites}b tend to converge to $\Phi_b=1$. When we increase $\rho_I$ to $0.6$, we observe that the variation between the different runs reduces dramatically while converging, and that all quartile of the distributions tend to converge to $\Phi_b=1$. Additionally, we can also notice that, with the increment of the proportion of informed robots from $\rho_I=0.1$ to $\rho_I=0.3$, the slope of the curve becomes slightly steeper during both the first and the second phase. That is, by progressively increasing $\rho_I$ the aggregation dynamics unfold in such a way that higher proportion of robots aggregated on the black site appear earlier during the run. To conclude, we can state that both speed (in terms of convergence) and accuracy (in terms of increase of percentage of robots aggregating on the desired site) of the aggregation process increase with increasing proportion of informed robots.

\subsection {Arena with three and four aggregation sites}
\label{sec::results:3_4_sites}
\begin{figure}[t]
	\begin{center}
		\begin{tabular}{cc}
			\includegraphics[width=0.28\textwidth]{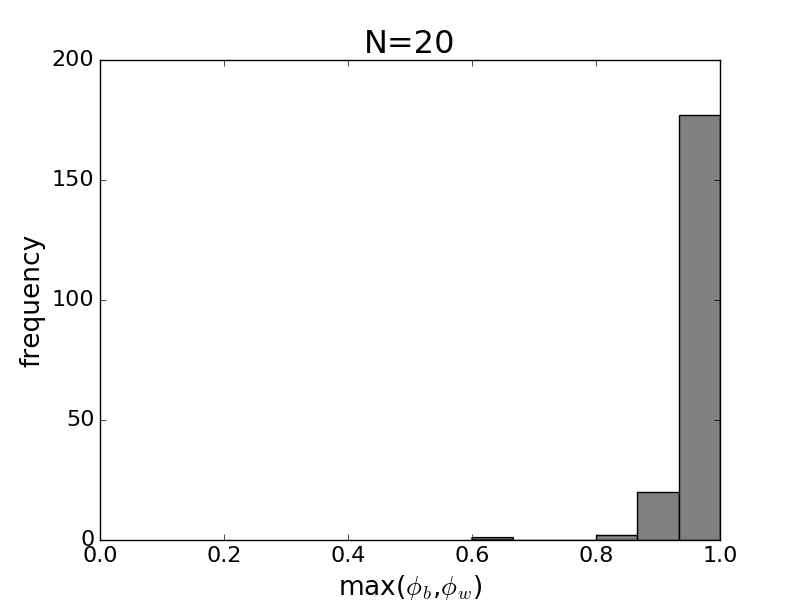} &
			\includegraphics[width=0.4\textwidth]{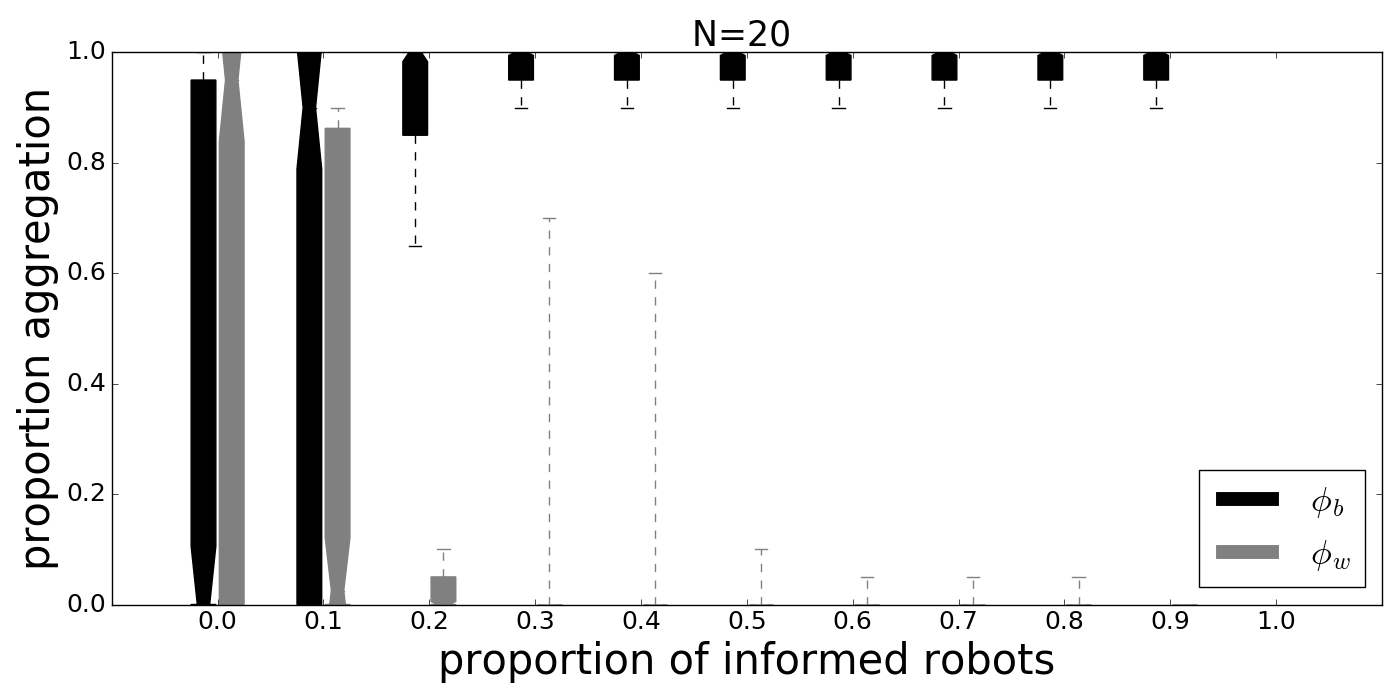} \\
			(a) & (b) \\
			\includegraphics[width=0.28\textwidth]{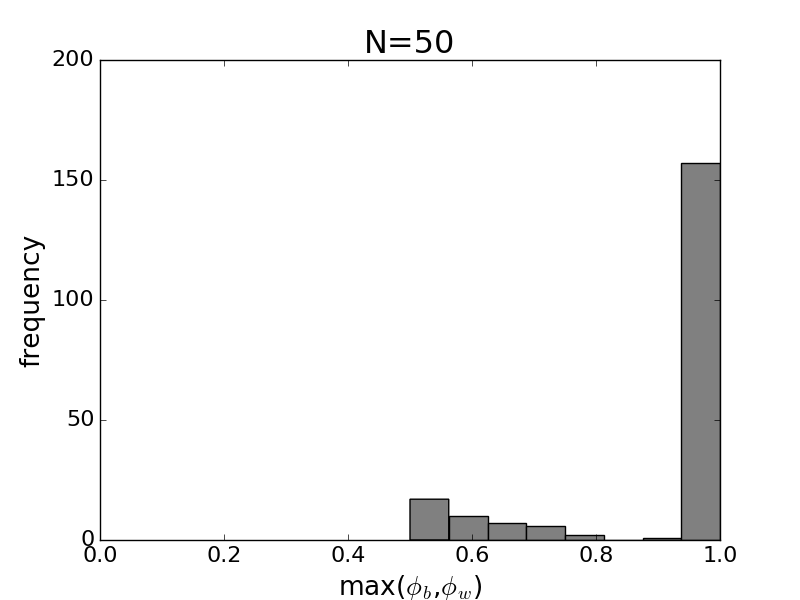} &
			\includegraphics[width=0.4\textwidth]{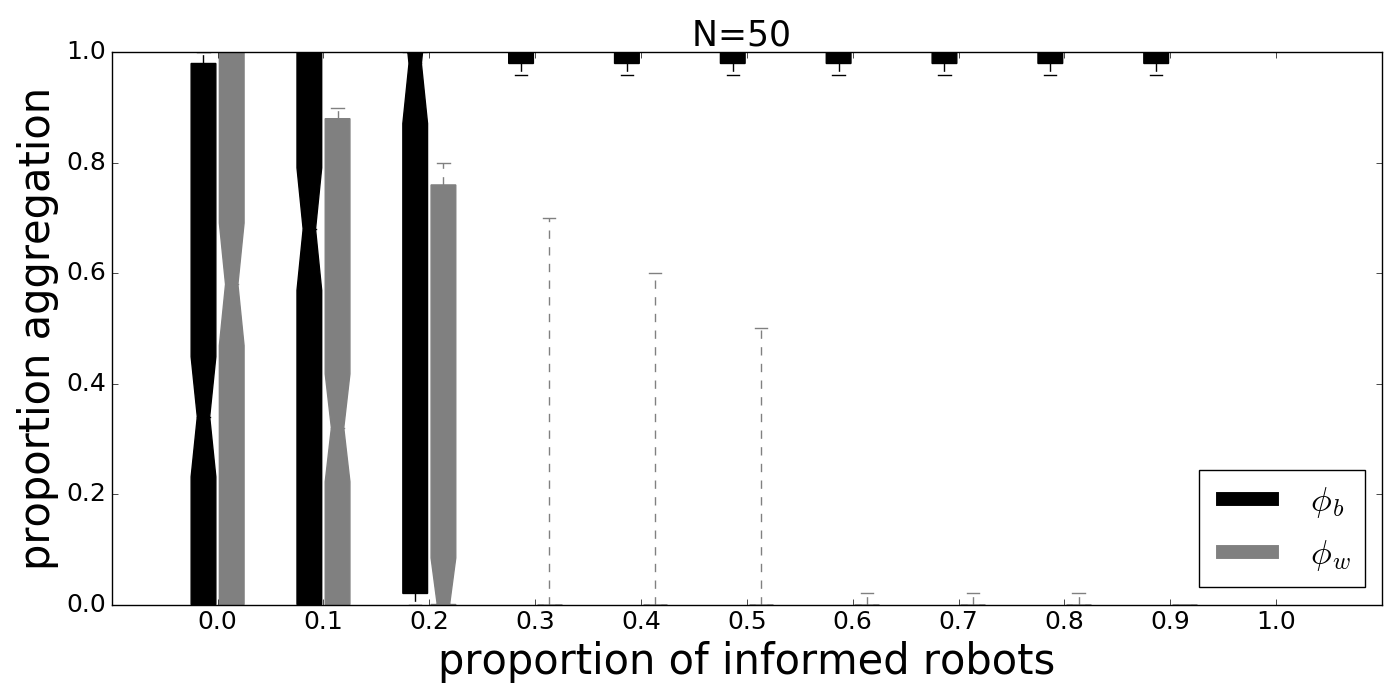}\\
			(c) & (d) \\
			\includegraphics[width=0.28\textwidth]{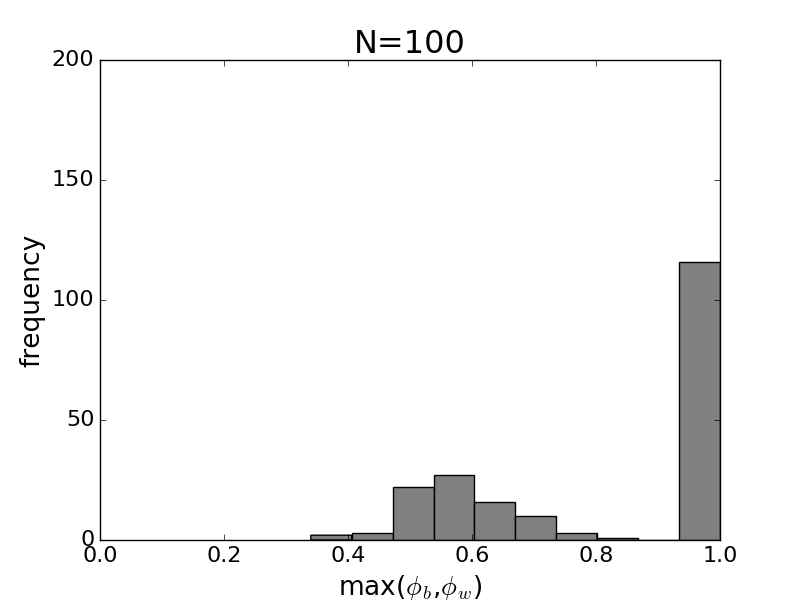}&
			\includegraphics[width=0.4\textwidth]{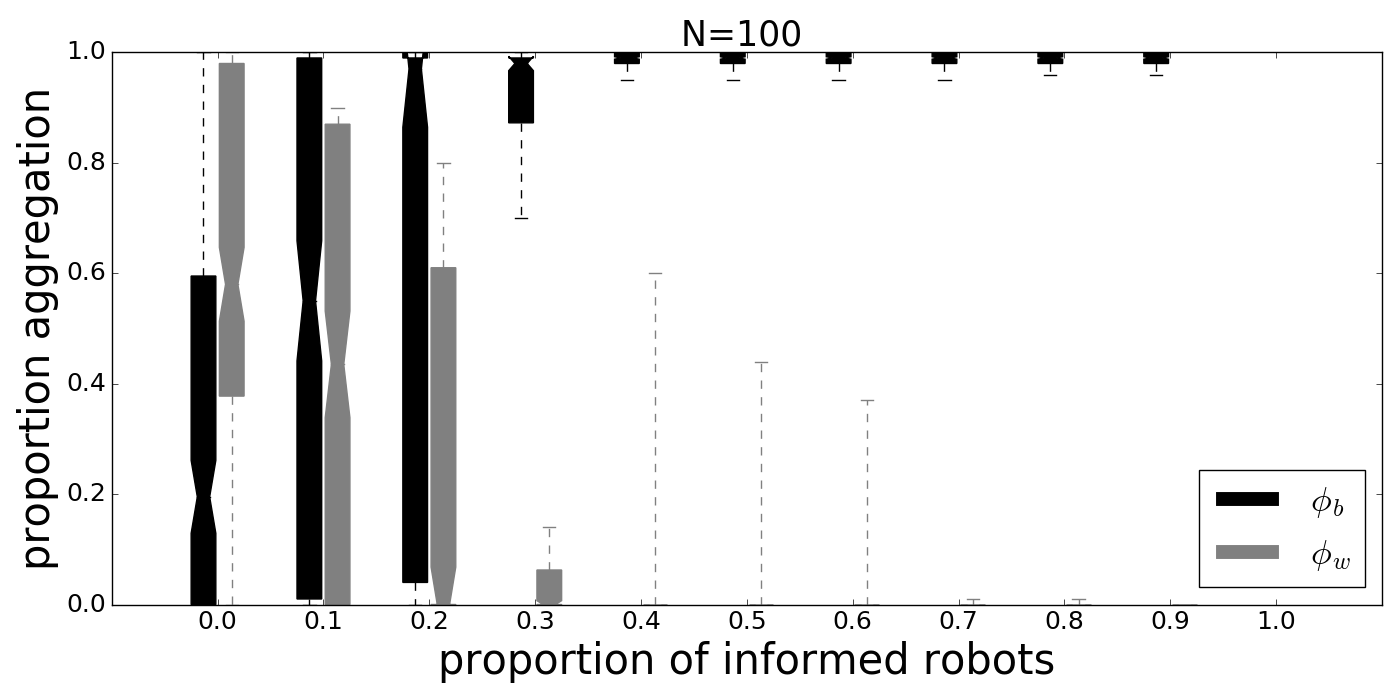}\\
			(e) & (f) \\
		\end{tabular}
		\caption{Results of the experiments in arena with three aggregation sites. The graphs in first column show frequency histograms of the proportion of robots aggregating on the largest aggregate ($\max(\Phi_b,\Phi_w)$) for swarms without informed robots ($\rho_I=0$) of size a) N=20; c) N=50; and e) N=100. Graphs on the second column show the percentage of aggregated robots on the white site (i.e., $\Phi_w$, see grey boxes) and on the black site ($\Phi_b$, see black boxes) for swarms of size b) N=20; d) N=50; and f) N=100. See also caption Figure~\ref{fig::2_sites} for more details.}
		\label{fig::3_sites}
	\end{center}
\end{figure}
In this section, we report the results of the simulations in a scenario with three sites (see Figure~\ref{fig::arena}b) and four sites (see Figure~\ref{fig::arena}c). 
 
The three-site scenario features an asymmetry in terms of the characteristics of the aggregation sites. White sites take up twice as much arena's area than the single black site. Thus, each robot is roughly twice as likely to find and eventually to stop on a white than on the black site. Our aim is to investigate whether and which proportion of informed robots is eventually required to invert the asymmetry and to induce the swarm to aggregate on the black site. As for the two-site scenario, prior to testing the effect of informed robots, we look at the frequency distribution of the largest aggregate for three different swarm sizes without informed robots ($\rho_I=0$). For this scenario, since it features a clear asymmetry in favour of the white site, we expect the swarm to systematically form large aggregate---with more than 90\% of the swarm---and to preferentially aggregate in any of the two white site. Results of this test are shown in Figure~\ref{fig::3_sites} first column. The graphs indicate that, independently of the swarm size, in the absence of informed robots, the distributions look multi-modal with the highest peak at 1.0. We have also observed that aggregates that include more than 90\% of the swarm's components occur 131 times on the white site and 66 times on the black site in 200 runs for N=20;  97 times on the white site and 70 times on the black site in 200 runs for N=50; 99 times on the white site and 41 times on the black site in 200 runs for N=100. This suggest that, without informed robots, large aggregates (i.e., aggregates with more that 90\% of the swarm's components) are relatively frequent (i.e., they occur in 98\% of the runs for N=20; in 84\% of the runs for N=50; in 70\% of the runs for N=100), and they are more likely to occur on a white than on the black site.

With the progressive introduction of informed robots ($0 \leq \rho_I \leq 1$)), the aggregation dynamics change quite radically as indicated in Figure~\ref{fig::3_sites} second column. For the small swarm size, $20\%$ of informed robots (i.e., $\rho_I=0.2$) is sufficient to invert the above mentioned trend, by generating a large majority of runs that end with more than $80\%$ of the robots aggregated on the single black site. For the medium and the large swarm size, a slightly higher proportion of informed robots (i.e., $\rho_I=0.3$) is required to observe the desired aggregation dynamics. As for the two-site scenario, also in the three-site scenario we observe that the higher the number of informed robots, the higher the proportion of robots aggregated on the black site. In summary, the above results indicate that with a proportion of informed robots varying from $0.2$ to $0.3$ of the entire swarm, it is possible to invert the swarm tendency to aggregate on the more represented (in terms of arena's area taken) type of site, and to generate robust and consistent aggregation dynamics that take the large majority of the swarm on the less represented (in terms of arena's area taken) type of site.

\begin{figure}[t]
	\begin{center}
		\begin{tabular}{cc}
			\includegraphics[width=0.28\textwidth]{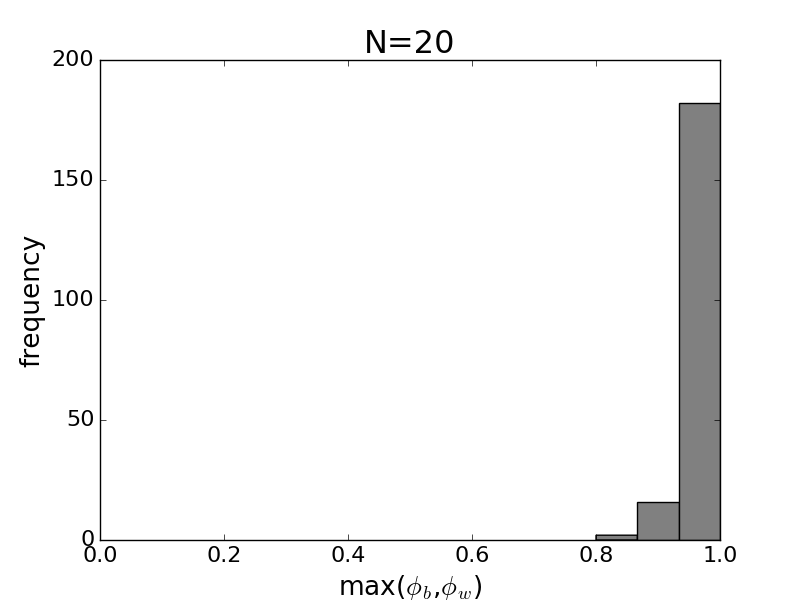} &
			\includegraphics[width=0.4\textwidth]{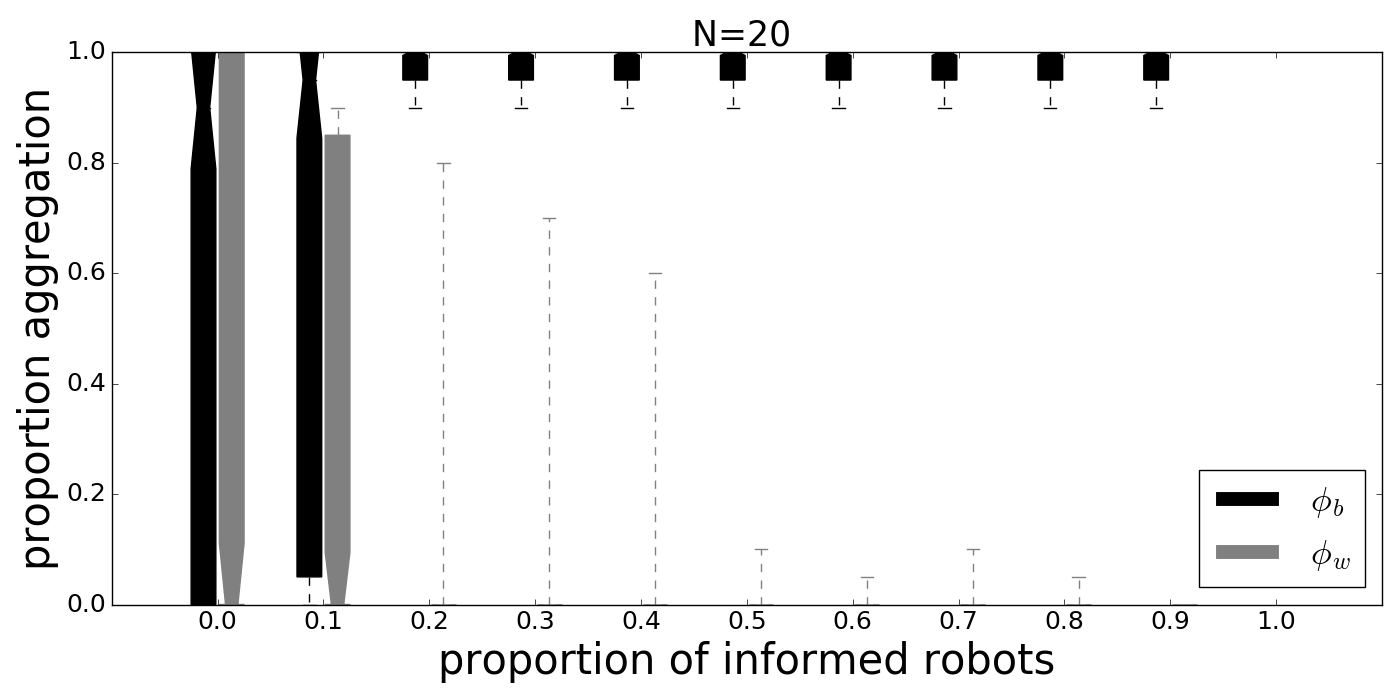} \\
			(a) & (b) \\
			\includegraphics[width=0.28\textwidth]{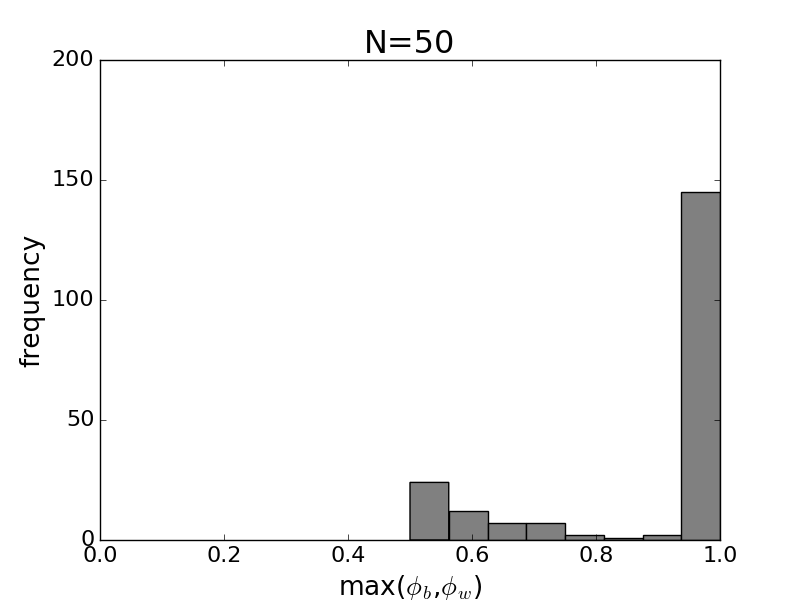} &
			\includegraphics[width=0.4\textwidth]{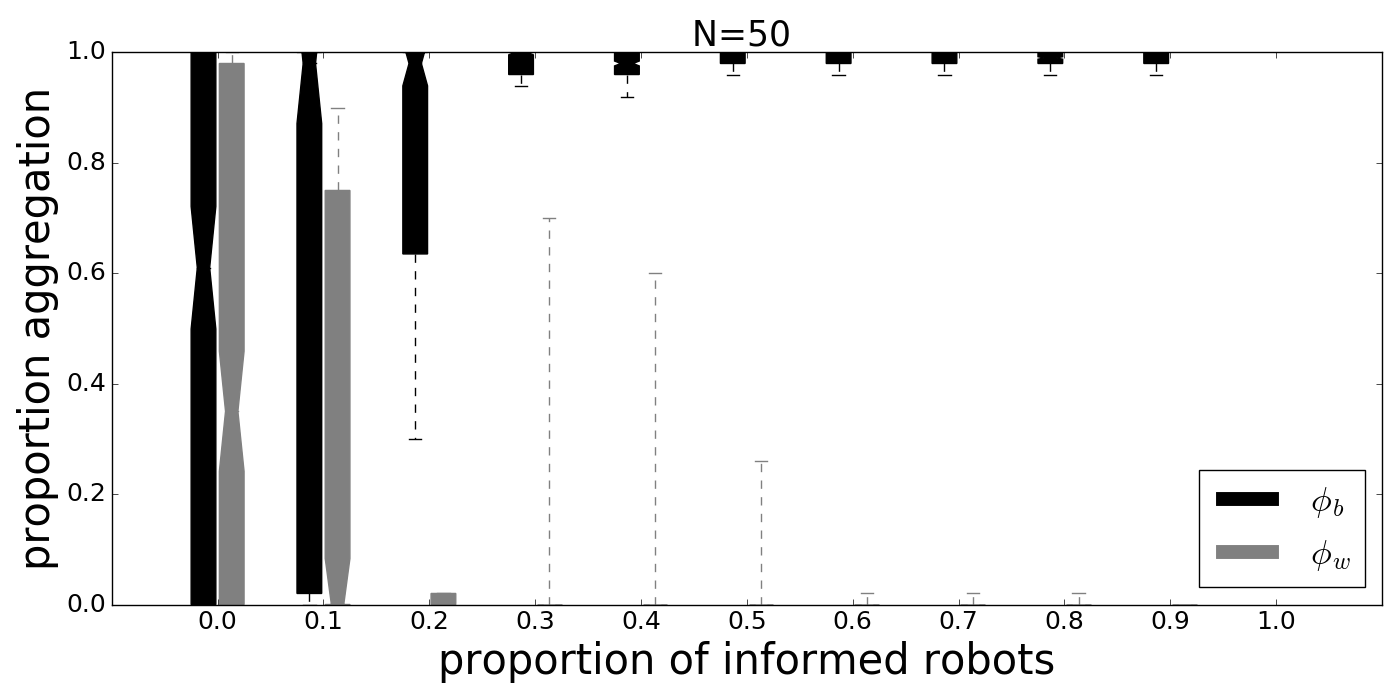}\\
			(c) & (d) \\
			\includegraphics[width=0.28\textwidth]{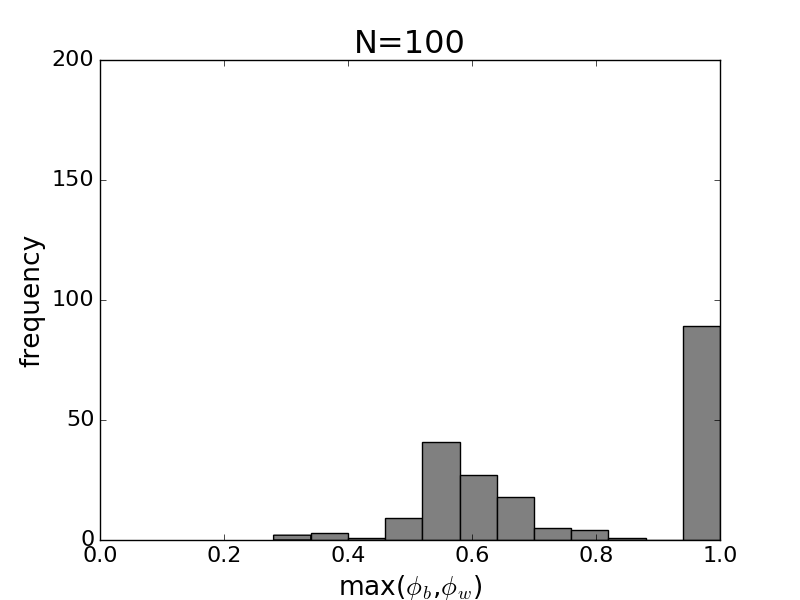} &
			\includegraphics[width=0.4\textwidth]{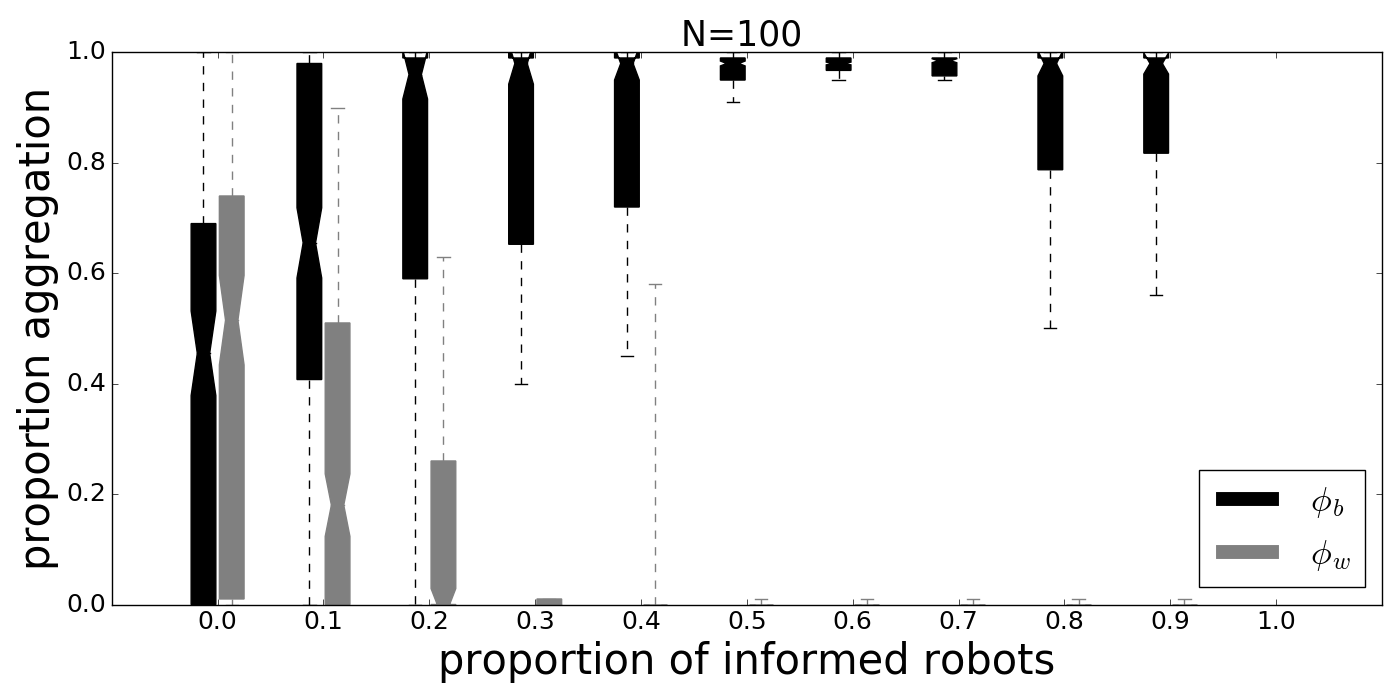} \\
			(e) & (f)\\			
		\end{tabular}
		\caption{Results of the experiments in arena with four aggregation sites. The graphs in first column show frequency histograms of the proportion of robots aggregating on the largest aggregate ($\max(\Phi_b,\Phi_w)$) for swarms without informed robots ($\rho_I=0$) of size a) N=20; c) N=50; and e) N=100. Graphs on the second column show the percentage of aggregated robots on the white site (i.e., $\Phi_w$, see grey boxes) and on the black site ($\Phi_b$, see black boxes) for swarms of size b) N=20; d) N=50; and f) N=100. See also caption Figure~\ref{fig::2_sites} for more details.}
		\label{fig::4_sites}
	\end{center}
\end{figure}
The four-site scenario, like the two-site scenario, is symmetric with respect to the arenas's area taken by the two types of site. However, the fact that there are two black sites instead of one represents a further challenge for the emergence of a single aggregate on a black site. Informed robots, which avoid to stop on white sites, are likely to stop on both black sites. This can be a deterrent to the formation of a single aggregate. For example, in the likely event in which informed and non-informed robots distribute on both black sites, a single aggregate can emerge only if the robots, including informed robots, in one of the two target sites leave that site for eventually joining the other target site.  Therefore, in this scenario more than in the two previously seen, it is the combination of the probability of staying on a site ($P_{stay}$) and the probability of leaving ($P_{leave}$) that generate the desired swarm dynamics.

As for the previous two experiments, we look at the frequency distribution of the largest aggregate for three different swarm sizes without informed robots ($\rho_I=0$). Since the scenario is symmetric, we expect the swarm to display the symmetry breaking property discussed above by forming large aggregate---with more than 90\% of the swarm---in any of the two types of site. Results of this test are shown in Figure~\ref{fig::4_sites} first column. The graphs indicate that, independently of the swarm size, in the absence of informed robots ($\rho_I=0.0$), the distributions look multi-modal with the highest peak at $1.0$. For $N=20$ and for $N=50$ large aggregate are relatively frequent and tend to occur in roughly the same quantity, on both types of site. For large swarms $N=100$ (see Figure~\ref{fig::4_sites}e) large aggregate are less frequent than for smaller swarms. The graphs in Figure~\ref{fig::4_sites}e shows that the highest peak at 1.0 occurs less than 100 times over 200 runs, with a quite frequent second highest peak at $0.5$ occurring about 50 times. This suggests that on this scenario, robots of large swarms are not as likely as robots of small and medium size swarms to form large aggregate (i.e., aggregates with more than 90\% of the swarm's components) on a single site. However, we observed that when the largest aggregate is larger than 90\% of the swarm components, the aggregate can be with about equal probability on one of the black or on one of the white sites. 

With the progressive introduction of informed robots ($0 \leq \rho_I \leq 1$), the aggregation dynamics change quite radically as indicated in Figure~\ref{fig::4_sites} second column. For the small and medium size swarm, $20\%$ of informed robots (i.e., $\rho_I=0.2$) is sufficient to result in the large majority of runs ending with more than $80\%$ of the robots aggregated on a single black site (see Figure~\ref{fig::4_sites}b, and~\ref{fig::4_sites}d). Moreover, the higher the proportion of informed robots, the higher the proportion of robots aggregated on a single black site. For the large swarm size, results are quite different, since the progressive increment of the proportion of informed robots does not result in a progressively higher proportion of robots aggregated on a single black site (see Figure~\ref{fig::4_sites}f). Observation of the behaviour of the simulated robots reveal that with more than 70\% informed robot (i.e., $\rho_I > 0.7$), swarms are very likely to form a large aggregate on a black site and a smaller aggregate on the other black site. In view of this, we claim that the aggregation dynamics generated by swarms with a high proportion of informed robots can be explained in the following. A high proportion (i.e., $\rho_I > 0.7$) of informed robots (i.e., robots that avoid to stop on white sites) in the swarm makes very likely the emergence of two aggregates, one on each black site. These aggregate can become large enough to generate cases in which each robot currently on black has enough neighbours to have an extremely low probability of leaving. Recall that the probability of leaving a site decreases with respect to the number of neighbours (see also section~\ref{sec::fsm}). This makes the two aggregates relatively stable. Therefore, anytime they emerge they are likely to last until the end of the run. This is also a consequence of the fact that the maximum number of neighbours a robot can perceived does not scale up with the swarm size, but it is bounded to 12 by the characteristics of the sensor used to collect this information. With progressively less informed robots these dynamics do not emerge even in large swarm. This is the reason why with a proportion of informed robots in $0.4 \leq \rho_I \leq 0.7$ we manage to systematically induce the swarm to aggregate on a single black site even in a scenario with two back and two white sites. This undesired effect observed in large swarms when $\rho_I > 0.7$ does not occur when the swarm size is smaller, since the neighbours of robots on black are rarely large enough to reduce the $P_{leave}$ of each single robot currently on a black site to the point at which any aggregate becomes stable. Thus, smaller aggregates on a black site tend to disappear relatively quickly. In summary, with a properly balanced proportion of informed robots, and irrespective of the swarm size, it is possible to generate robust and consistent aggregation dynamics that take more than 90\% of the swarm's components on a single black site in a scenario with two black and two white sites and also in an asymmetric scenario (i.e., two white and one black site) that tends to favour aggregation on the undesired white site.

\section {Analysis of aggregation dynamics with an ODEs' model}
\label{sec::ode_model}
\begin{figure}[t]
	\begin{center}
		\begin{tabular}{ccc}
			\includegraphics[width=0.31\textwidth]{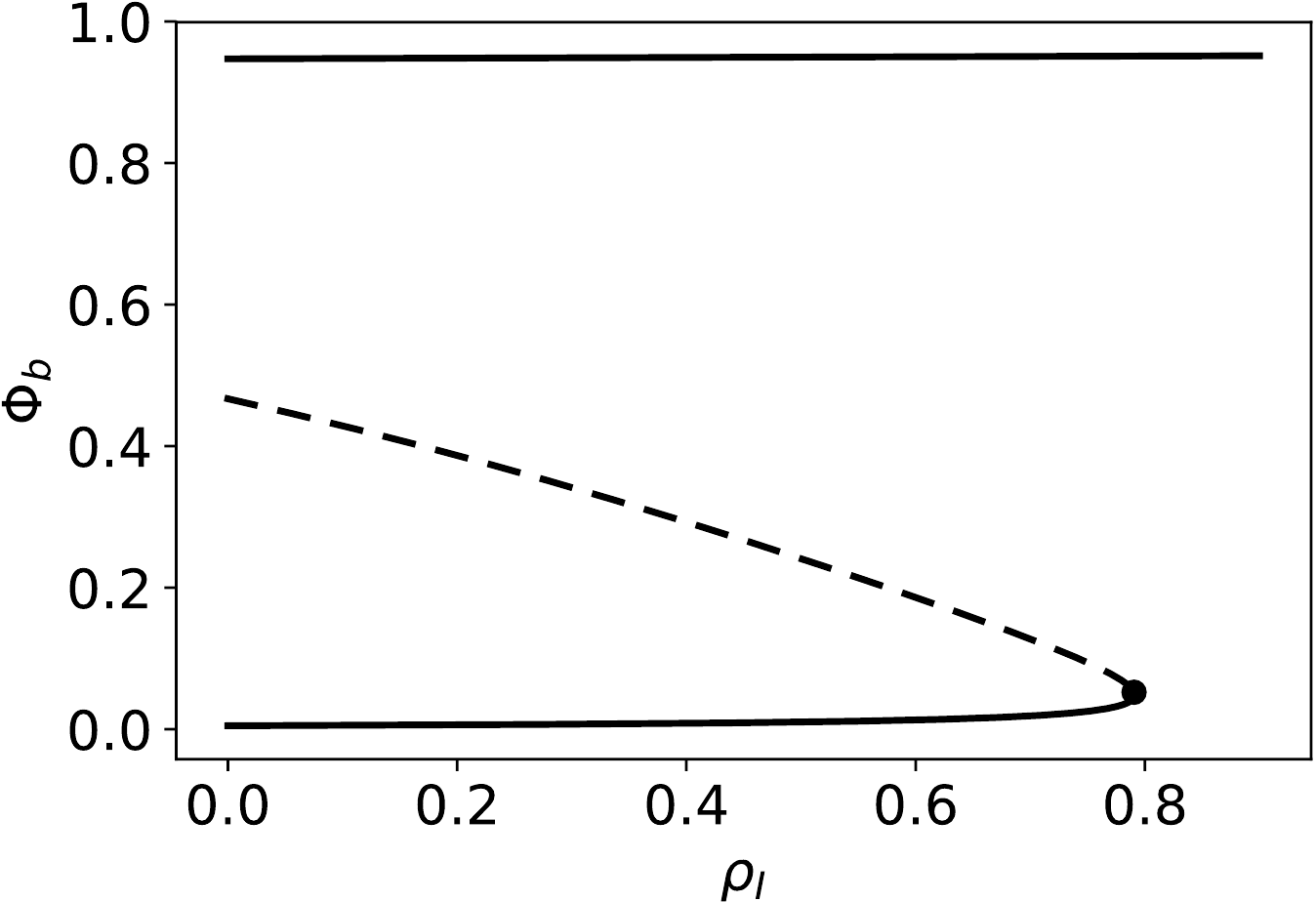}& 
			\includegraphics[width=0.31\textwidth]{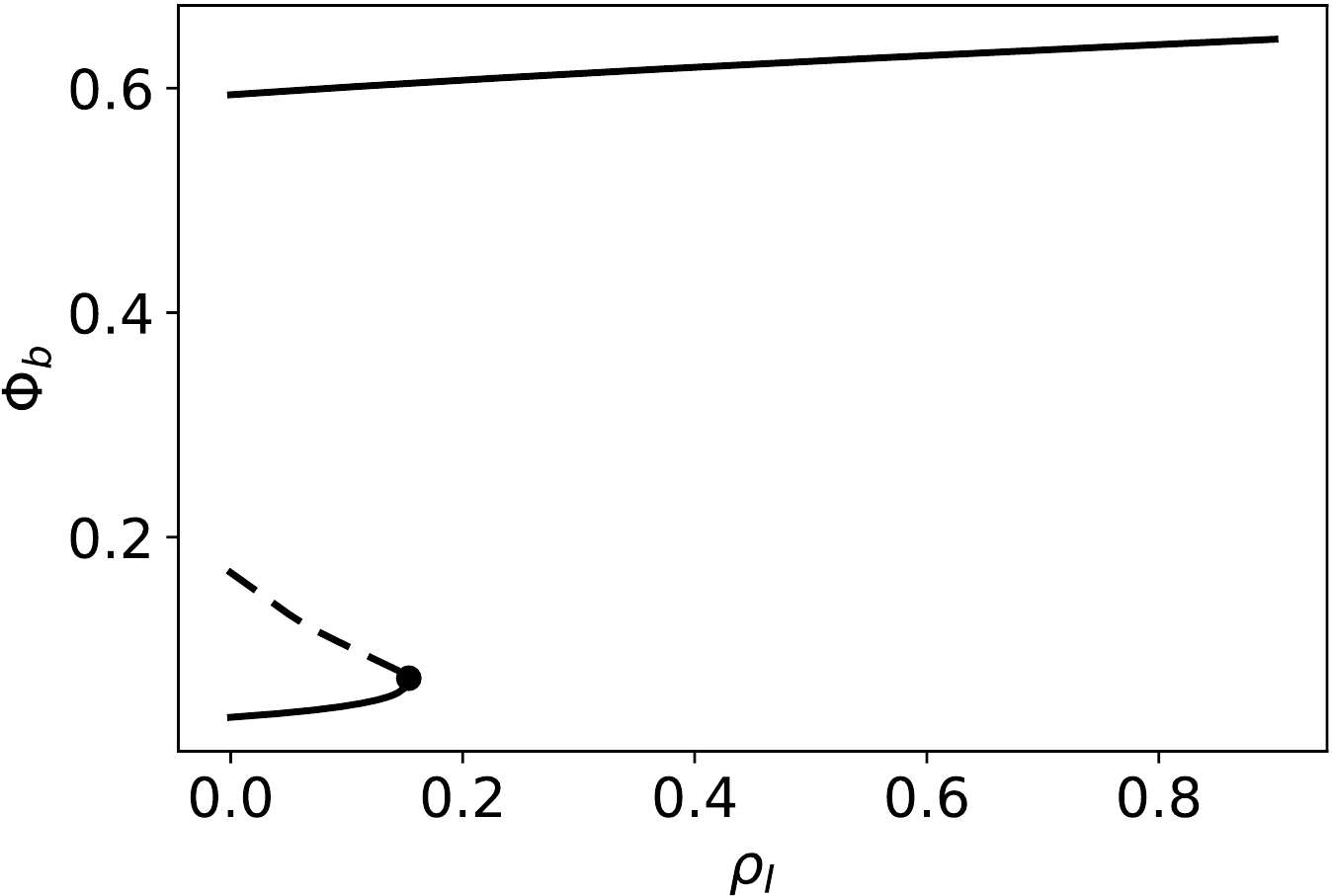}&
			\includegraphics[width=0.31\textwidth]{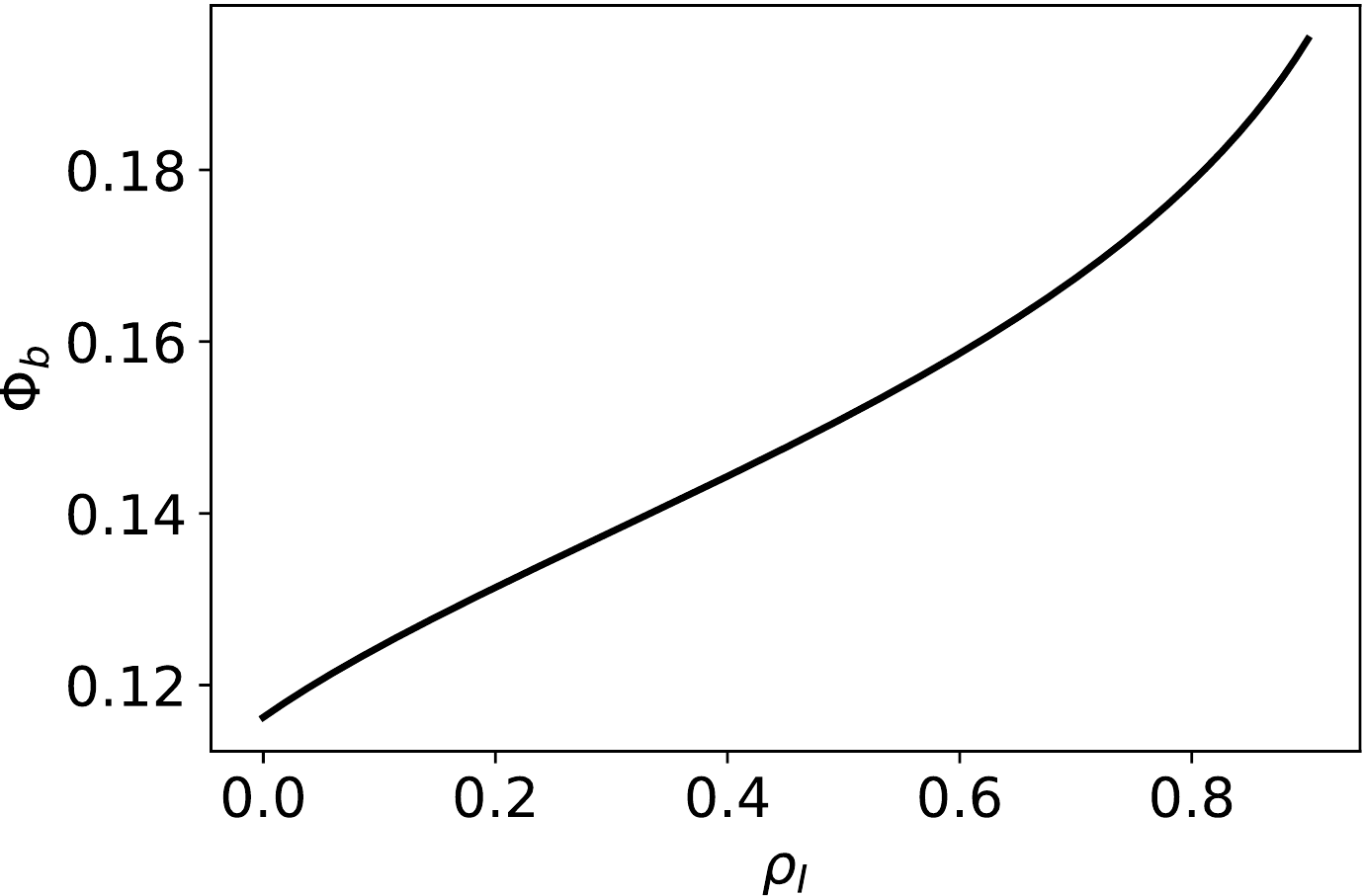}\\
			(a) & (b) & (c) \\
			\includegraphics[width=0.31\textwidth]{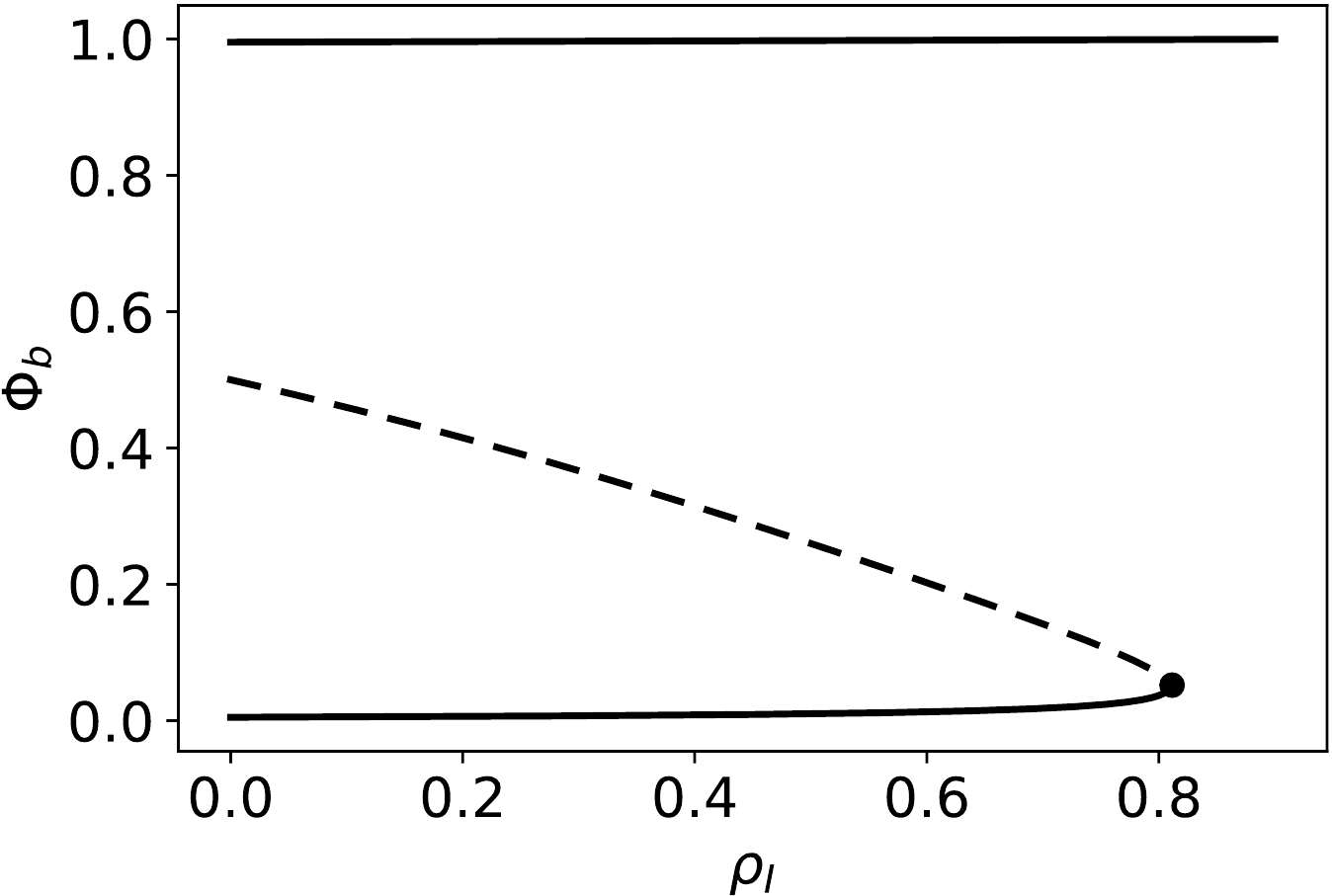}& 
			\includegraphics[width=0.31\textwidth]{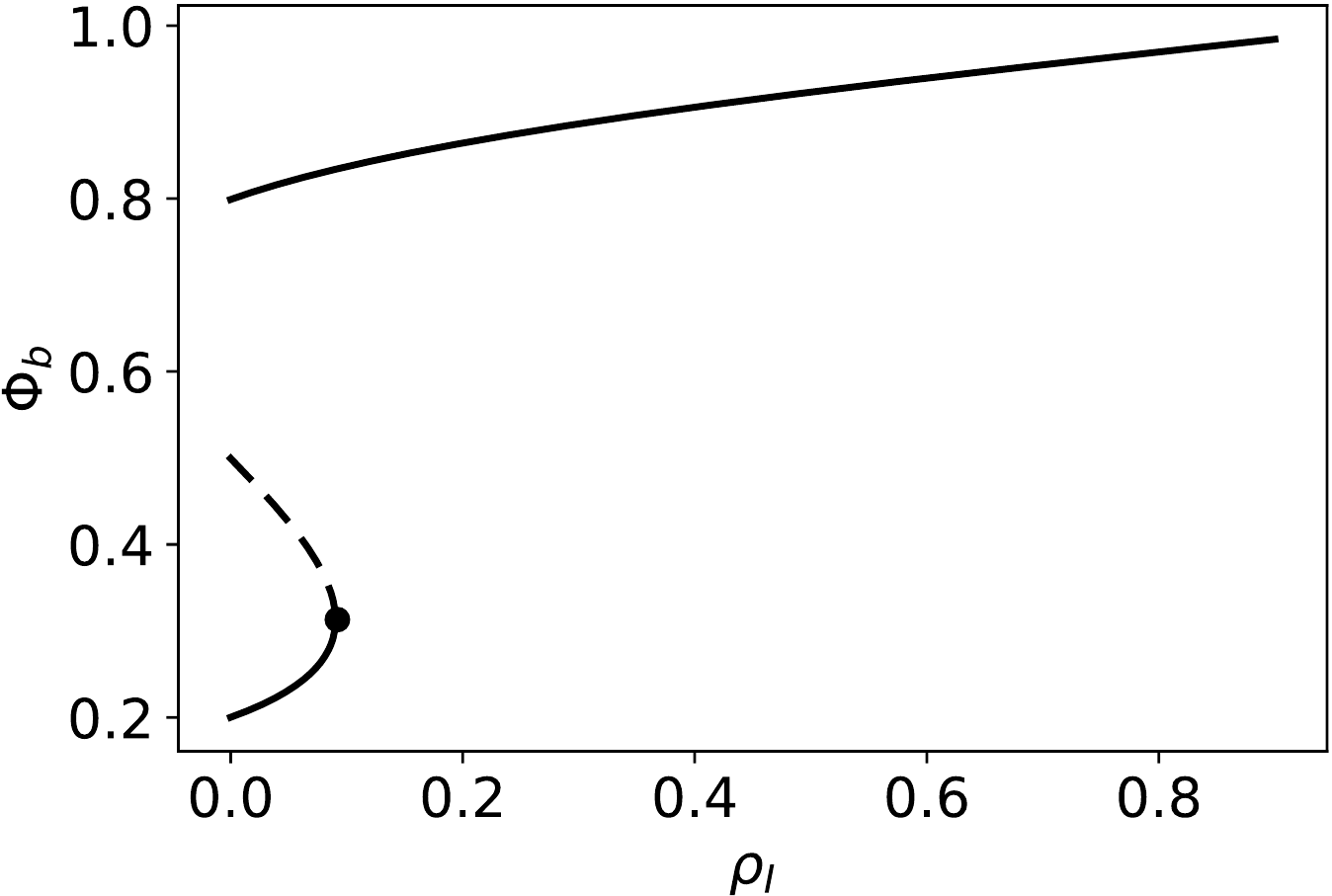}&
			\includegraphics[width=0.31\textwidth]{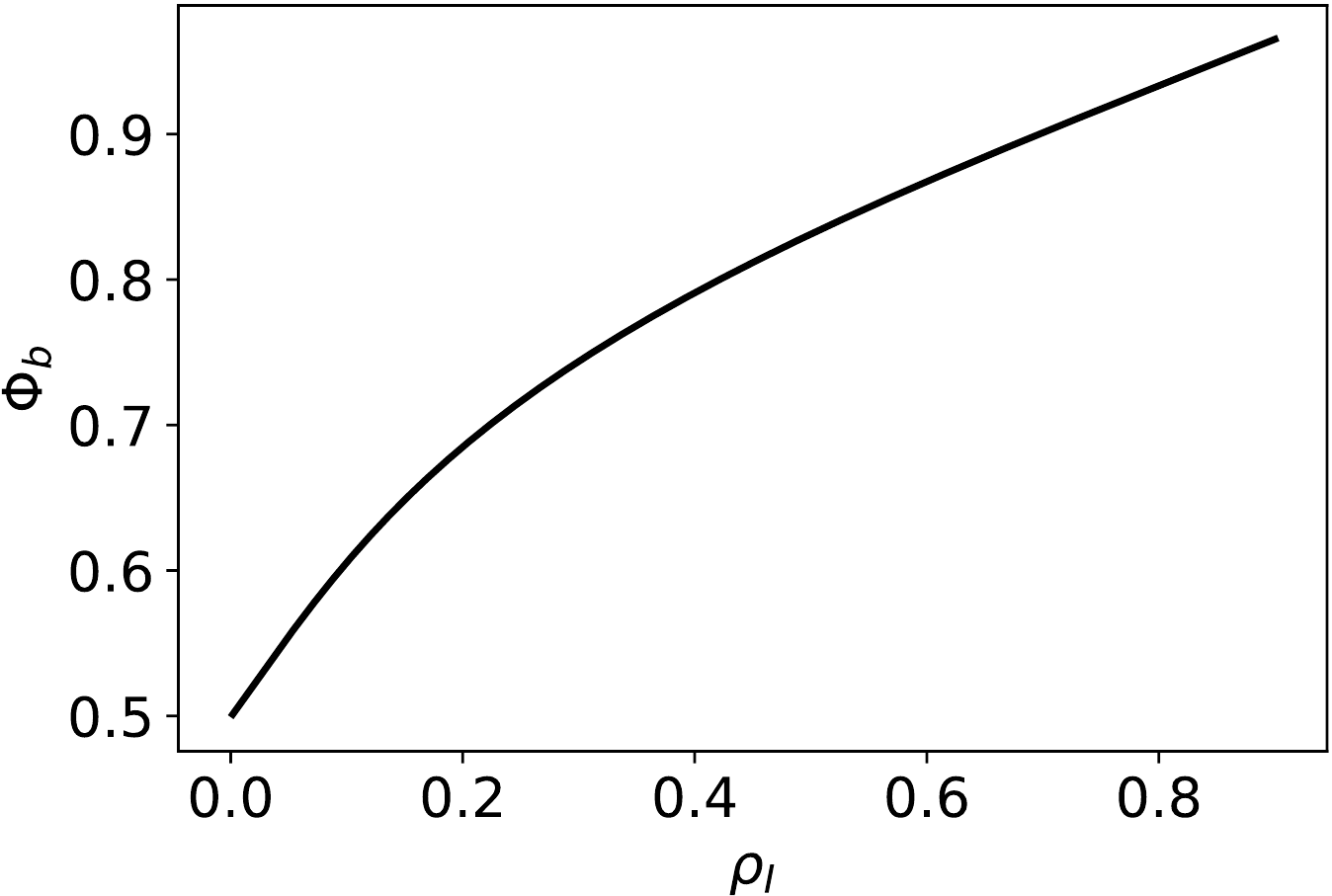}\\
			(d) & (e) & (f) \\
		\end{tabular}
		\caption{Bifurcation diagrams of the system of ODEs we propose as an extension of the model in~\citep[][]{AmeEtAlPNAS2006}. These plots show what happens to the proportion of robots aggregating to the black site $\Phi_b$ as a function of the proportion of informed robots $\rho_I$. Continuous line represent stable equilibria, while dashed lines represent unstable ones (i.e. saddle points). The big filled dots represent bifurcation points, in this case saddle-node bifurcations. In all plots, $\epsilon = 0.01$. In the first row, $\mu=0.001$, while in the second row $\mu = 1.5$. Parameter  $\gamma$ is set to the following values. (a) $\gamma=1667$, (b) $\gamma=250$, (c) $\gamma=180$, (d) $\gamma=1667$, (e) $\gamma=50$, (f) $\gamma=20$. For all graphs, $\sigma=\frac{S}{N}=2.0$.}
		\label{fig::ame::replica}
	\end{center}
\end{figure}
In this section, we complement our study on the aggregation dynamics using a macroscopic mathematical model. The model we propose is based on ODEs and is targeted at studying the two-site scenario (see Figure~\ref{fig::arena}a). In particular, we extend the ODEs originally introduced and discussed in~\citep[][]{AmeEtAl2004_AB} and subsequently reused in other studies on self-organised aggregation~\citep[e.g.,][]{AmeEtAlPNAS2006,CampoEtAl2010}. 
The extension we propose includes a term that models the presence of the informed robots. The resulting model is the following:
\begin{empheq}[left=\empheqlbrace]{align} 
\dot{N_b} \hspace{-0.0cm}& = -N_{b}\lambda_{b} + \mu \Big(1-\frac{N_{b}}{S} \Big)\Omega 
;\label{eq::ode_b}\\
\dot{N_w} \hspace{-0.0cm}& = -N_{w}\lambda_{w} + \mu \Big(1-\frac{N_{w}}{S} \Big)\Omega (1-\rho_I);\label{eq::ode_w}
\end{empheq} 
with
\begin{eqnarray}
\lambda_{b} &=& \frac{\epsilon}{1 + \gamma \Big(\frac{N_b}{S}\Big)^{2}}; \qquad \lambda_{w} = \frac{\epsilon}{1 + \gamma \Big(\frac{N_w}{S}\Big)^{2}};\label{eq::lambda}
\end{eqnarray}
and
\begin{eqnarray}
\Omega &=& (N-N_{b}-N_{w});\label{eq::omega}
\end{eqnarray}
In this model, the two state variables are the number of robots on the black and white sites, that is $N_b$ and $N_w$. $\mu$ represents the ``discovery rate'' of a site, $\lambda_{b}$ is the rate of leaving a black site, $\lambda_{w}$ is the rate of leaving a white site, $N$ is the swarm size, and $S$ is the site carrying capacity. The leave probabilities for the black ($\lambda_{b}$) and the white site ($\lambda_{w}$) are expressed as sigmoidal response to the density of robots ($\frac{N_{b/w}}{S}$) at each site, with $\epsilon$ and $\gamma$ being the parameters of the sigmoid. The probability to join a site is $(1-\frac{N_{w/b}}{S})$, which is $1$ when no other robots is on the site, decreases linearly as the number of robots on the site increases, and is $0$ when the site is full. The number of robots on the black (resp. white) site decreases proportionally to the leave probability $\lambda_{b/w}$ and to the current number of robots on the black (resp. white) site. It increases proportionally to the rate $\mu$ at which a site is encountered, to the join probability $(1-\frac{N_{w/b}}{S})$, and to the proportion of robots ``free roaming'' available to join a site, that is robots that are in neither of the aggregates $\Omega$. In our version of the model, we also introduce the term $(1-\rho_I)$ for the equation concerning the white site ($w$). This term rescales the number of robots that can potentially join a white site to only the non-informed robots, since informed robots never rest on the white site. The model in~\citep[][]{AmeEtAlPNAS2006} can be recovered by setting $\rho_I=0.0$.  

In the original study~\citep[][]{AmeEtAlPNAS2006}, some model parameters were kept fixed while others were studied in details. The parameters that were kept fixed were tuned after the experiments performed with the real cockroaches, and the corresponding values were: $\mu=0.001 s^{-1}$,  $\epsilon = 0.01 s^{-1}$, and $\gamma = 1667$. Different values for the ratio $\sigma=\frac{S}{N}$ were tested. Only when $\sigma > 1$ ($S>N$) (e.g., when $N=100$, and $S=200$) the swarm manage to fully aggregate on a site. In this study, we will only consider the case $S>N$, but we will study how the dynamics change by varying the other parameters $\mu$, $\epsilon$, and $\gamma$, as those are the parameters that vary when robotics experimental conditions (such as the size of the arena, the size and speed of the robots, the random walk strategy followed, etc \dots ) are varied.

Our main objective is to  study the effects of different proportions of informed robots ($0 \leq \rho_I \leq 1$) on the aggregation dynamics, and in particular on the proportion of robots on the black site ($\Phi_b$). This analysis is best exemplified using a bifurcation analysis, that is, by checking what happens to the steady states of the systems when we vary our key parameter $\rho_I$. For the sake of generality, we report on the $y$-axis the proportion, rather than the number, of robots on the black site $\Phi_b=\frac{N_b}{N}$.

To start with, we consider the original values of the parameters studied in~\citep[][]{AmeEtAlPNAS2006}, that is $\mu=0.001 s^{-1}$,  $\epsilon = 0.01 s^{-1}$, $\gamma = 1667$, and $\sigma=2.0$. As shown in Figure~\ref{fig::ame::replica}a the model predicts that when $\rho_I$  is smaller than a critical value very close to $0.8$, two equilibria exist: $\Phi_b \approx 1$ and $\Phi_b \approx 0$. This means that the two states represented by having all robots aggregated to the black or to the white site are equally likely, according to the model. This result is exactly the same obtained in~\citep[][]{AmeEtAlPNAS2006}. However, at the critical threshold for $\rho_I$, a saddle-node bifurcation occurs. After this value, only one stable equilibrium is found, which corresponds to the state seeing all robots aggregated on the black site. In other words, the model predicts that, in the original experimental setup with cockroaches, about $80\%$ of the cockroaches would need to be informed in order to have aggregation on the website.

Since our robots are not cockroaches, we were curious to see what happened when varying the parameters, in order to get as close as possible to the regime observed in the robotics simulations. We observed that by varying the parameter $\epsilon$ and $\sigma$ no changes is observed with respect to the proportion of robots required to get the entire swarm aggregate on the black site. Instead, we observed interesting changes at the variation of the parameter $\gamma$. As shown in Figure~\ref{fig::ame::replica}b, with a much lower value  of $\gamma=250$,  the bifurcation occurs much earlier in the $\rho_I$ parameter space. However, the higher stable equilibrium is no longer close to $1$, but to $0.6$. This means that slightly more than half of the robots can aggregate on black site with as little as 20\% of informed robots, but there is no way to have all robots aggregating on the black site. When we further decrease  $\gamma$ to $180$, we observe completely different dynamics, whereby the bifurcation no longer exists and even smaller aggregates form, with a size that increases with increasing $\rho_I$, reaching a maximum of about $20\%$ of the swarm (see Figure~\ref{fig::ame::replica}c). 

In the above analysis, we observed that none of those parameters could replicate the results that we obtain in simulation, that sees robots aggregating in large proportions to the black site with a small proportion of informed robots. Therefore, we hypothesised that the remaining parameter, $\mu$, had to be studied in order to find a regime in which our simulation experimental results could qualitatively be reproduced. After exploring this new parameter, we found that those dynamics can be reproduced with a much larger value for the site discovery rate $\mu$.

We report in the bottom row of Figure~\ref{fig::ame::replica} the results of the analysis performed with $\mu=1.5$ and with $\gamma=1667$ (Figure~\ref{fig::ame::replica}d), $\gamma=50$ (Figure~\ref{fig::ame::replica}e), and $\gamma=20$ (Figure~\ref{fig::ame::replica}f). As we can see, the results are qualitatively similar to those obtained with low discovery rate $\mu=0.001 s^{-1}$, but the stable state in the higher branch is now very close to $1$. When $\gamma=1667$, the critical ratio of informed robots is above $80\%$. When $\gamma=50$, we obtain near ideal results, similar to those obtained in simulation, whereby the critical ratio of informed robots is below $20\%$. With $\gamma = 20$ once again no bifurcation occurs, however with $100\%$ informed robots the system is able to aggregate all robots on the black site.

Although the fine tuning of the model parameters after the simulation is out of scope for this paper, we have clearly seen here for the first time that the extended model of~\citep[][]{AmeEtAlPNAS2006} is indeed very rich in terms of dynamics. Experimental parameters $\mu$ and $\gamma$ play a crucial effect. However, many of these are out of the designer controls, with the exception of $\gamma$, a component of the leave probability. This suggests that the leave probability is of critical importance when designing self-organized aggregation, even in presence of informed robots.

\section{Conclusions}
\label{sec::conclusions}
In this paper, we have contributed to the wider agenda of studying the role of implicit leaders in the context of collective decision making in swarms of robots. We have focused on self-organised aggregation in scenarios that feature two types of aggregation sites: a site coloured in black and a site coloured in white. The circular arena's floor where the robots operate is coloured in grey. We studied three different scenarios: two symmetrical scenarios, in which either two or four aggregation sites are available in the environment, and an asymmetrical scenario in which three aggregation sites are available. In the symmetric scenario, the two types of aggregation site (the black and the white) are equally represented. In the asymmetric scenario, one type of aggregation site (the white) is twice as much represented than the other type of site. In all scenarios, the robots are required to form a single aggregate, on a black site. We considered a swarm of robots divided in two sets: informed robots, that possess extra information on which site the swarm has to aggregate. Therefore, they selectively avoid to stop on any white site. Non-informed robots do not possess this extra information. Therefore, they are equally likely to stop on a white and on a black site according to the mechanisms of the finite state machine that controls their behaviour. The objective of this study is to look at whether and eventually which proportion of informed robots is required to direct the aggregation process toward a pre-defined type of site (i.e., the black) among those available in the environment.

We conducted experiments using the ARGoS simulator in which we varied the proportion of informed robots from 0\% to 100\%. Our results show that, in absence of informed robots, in all the three different scenarios, and for different swarm size, the swarms tend to form a single large aggregate (i.e., aggregates made of more than 90\% of the robots). In symmetric scenarios, these large aggregates emerge with almost equal frequency on both the black and the white site. In the asymmetrical scenario, the large aggregates are more frequently observed on the most represented type of site (i.e., the white one). The original contribution of this study is in showing that the above mentioned dynamics can be modified with as little as 20\% of informed robots. In particular, in the simplest two-site scenario, we show that when at least $20\%$ of the robots are informed, the entire swarm aggregates on the black site, for all swarm sizes we have considered. We have also shown that the speed and accuracy of convergence is also strongly affected by the proportion of informed robots. In the asymmetrical three-site scenario, largest aggregates can be easily induced to emerge entirely on the black site with 20\% to 30\% of informed robots depending on the swarm size. On the four-site symmetrical scenario, 20\% of informed robots are sufficient to systematically generate large aggregates on one of the black site. However, for large swarms (i.e., swarm size $N=100$) the presence of too many informed robots in the swarm is counterproductive, since it frequently leads to the formation of more than one aggregate on different black sites. We believe that the formation of multiple aggregates in this scenario is a results of the relationships between the high proportion of informed robots, the robots' perceptual apparatus used to detect and count neighbouring agents currently resting on a site, and the mechanisms that regulate the probability of a robot to leave a site. Future work, in which we will explore the relationships among these three factors, by systematically varying them, are needed to fully corroborate our claim.

Another valuable contribution of this study is the analysis of the ODEs model discussed in~\citep[][]{AmeEtAlPNAS2006} to account for the dynamics of self-organised aggregation observed in cockroaches by calling upon the principle of attraction between individuals. In~\citep[][]{AmeEtAlPNAS2006} and in~\citep[][]{CampoEtAl2010}, this model is used to investigate how the aggregation dynamics changes by varying the size of aggregation site and consequently their carrying capacity. We extended the model by introducing the concept of informed robots. By exploring the parameters of the model, we show that under specific conditions, the model predicts the results we observed in the simplest two-site scenario (see section~\ref{fig::2_sites}). That is, the model shows that with about 20\% of informed robots the emergence of a large aggregate on black is a stable equilibrium of the system. The analysis of the model's parameters leads to a deeper understanding of the relationships between environmental features and agents' exploration strategies. We show how these relationships bear upon the emergence of a single aggregate and how they interfere by amplifying or by reducing the effects of informed robots on the group aggregation process.

This study has the potential to be extended in many possible ways. First, in the context of aggregation, our next step will be to extend the study to more complex scenarios. We plan to test the discrimination capabilities of our swarms with informed robots in environments with several different options (e.g. colours), which would correspond to a best-of-$n$ problem with $n>2$~\citep[][]{Valentini2017Review}; scenarios where informed robots may have conflicting information about which is the best site and conflict resolution strategies need to be devised. Secondly, in our vision, we also plan to introduce implicit leaders in other collective behaviours. Our framework can also have a practical relevance in the context of human-swarm interaction~\citep[see][]{SycaraHumanSwarm2016}, whereby informed robots can correspond to robots that are controlled or tele-operated by humans, which would in turn introduce the human in the loop in order to study how humans can interact and control swarms of robots. Finally, we intend to generate an ODEs' model based on the mechanisms of the finite state machine controller as illustrated in section~\ref{fig::controller}. The model would facilitate the investigation of the effects of informed robots on the aggregation dynamics by varying the parameters that regulate the probability of joining/leaving a site, the type of exploration strategies (e.g., the type of random walk) used to search for the aggregation site, and by varying the swarm density in the arena.

\section*{Conflict of Interest}
The authors declare that they have no conflict of interest.
 


\end{document}